\documentclass[a4paper,11pt]{article}
\pdfoutput=1

\usepackage{jheppub}
\usepackage{xcolor}
\usepackage[T1]{fontenc}
\usepackage{xspace}
\usepackage{multirow}
\usepackage{hyperref}
\usepackage{slashed}
\usepackage{amsmath}
\usepackage{mathtools,amssymb,mathrsfs,dsfont}
\usepackage[utf8]{inputenc}
\usepackage{tikz}
\usetikzlibrary{decorations.pathreplacing}
\usepackage{subfigure}
\newcommand{\Z}{{\mathds Z}}
\newcommand{\one}{{\mathds 1}}
\newcommand{\Lqcd}{\Lambda_{QCD}}
\newcommand{\br}{{\rm br}}

\def\GeV{\ifmmode {\mathrm{\ Ge\kern -0.1em V}}\else \textrm{Ge\kern -0.1em V}\fi}%

\preprint{IPPP/26/82}

\title{\boldmath Monopole catalysis with massive fermions: the fate of the twisted sector}

\author{Valentin V. Khoze}

\medskip
\medskip

\affiliation{Institute for Particle Physics Phenomenology, Department of Physics, Durham University, \newline Durham DH1~3LE, U.K.}

\bigskip

\emailAdd{valya.khoze@durham.ac.uk}

\abstract{In the massless limit a fermion scattering on the minimal $SU(5)$ monopole in the lowest partial wave produces an outgoing state with half-integer fermion number in every flavour. This state lives in a twisted sector of the Hilbert space and is now understood as an ordinary fermion attached to a topological line running back to the monopole. In the Standard Model fermion masses and confinement destroy the line at distances greater than the inverse mass scales, causing the twisted state to decay into physical asymptotic states with integer fermion numbers. We study this decay for an incoming lepton or quark of one generation. The cross-section in this partial wave saturates the unitarity bound, $\sigma=\pi/p^2$, and the energy of the outgoing state is shared among the four fermion flavours with a broad Dirichlet distribution, which sets the thresholds and the kinematics. Conservation laws allow exactly two decay channels for each incoming fermion, the helicity flip and its $(B+L)$-violating partner. For an incoming electron the $(B+L)$-violating channel is anti-nucleon synthesis, and we determine the branching ratios in two limits: the two channels are equally likely at high energy, and only the helicity flip remains at low energy, below the anti-nucleon threshold. In between we model the energy dependence. For a quark in a slow proton the $(B+L)$-violating channel is catalysed proton decay, which has no threshold and occurs with a probability close to one. Callan's statistical interpretation of the fractional state emerges in a sharper form: the state is not a superposition of two final states with equal weights, but a single state that decays into them with energy-dependent probabilities, equal only at high energy.}

\begin{document}
\maketitle

\tableofcontents

\flushbottom


\section{\label{Sec:Intro}Introduction and summary of results}

When a proton collides head-on with a magnetic monopole, the proton is destroyed and its baryon number is converted into minus the lepton number of the final state. The amplitudes are not suppressed by inverse powers of the unification scale or of the monopole mass, nor by any semiclassical exponent: the $(B+L)$-violating reactions saturate unitarity in the lowest partial wave. This is the Callan--Rubakov phenomenon of monopole catalysis of proton decay~\cite{Rubakov:1981rg,Rubakov:1982fp,Callan:1982ah,Callan:1982au,Callan:1982ac},
$p\,+\,  M \,\to\, M\,+\, \overline{e}\,+\, X$. The process is inclusive and  $X$ denotes unspecified $(B+L)$-preserving radiation (photons, mesons, lepton pairs).
At the partonic level, for the minimal $SU(5)$ monopole $M$, the underlying reaction is\footnote{To keep the presentation simple, here we consider only the minimally-charged monopole that we denote $M$. Higher monopoles, $M_{p}$ and $M'_p$, and Callan--Rubakov scattering of Standard Model fermions on them were studied in~\cite{Khoze:2024hlb}.}
\begin{equation}
 u_L^{\,1} \,+\,u_L^{\,2} \,+\,  M \,\to\, M\,+\,(\overline{d}\,)_{R3} \,+\, (\overline{e})_{R}\,+\, X\,,
\label{eq:m-cat-p}
\end{equation}
where two valence quarks of the proton scatter on the monopole in the lowest partial wave, the $j_0$-wave,\footnote{We use the $s$-wave and the $j_0$-wave interchangeably. Both refer to the lowest partial wave in the fermion--monopole scattering. An infinitely heavy monopole is frozen and plays the role of the boundary at the origin on which  $s$-wave fermions scatter.} in which there is no centrifugal barrier and the incoming fermions reach the monopole core~\cite{Kazama:1976fm}.

The problem becomes more subtle for a single incoming fermion, say a right-handed electron. An electron in the $j_0$-wave also reaches the monopole core, and in the massless limit the Callan--Rubakov boundary condition produces an outgoing state with fractional fermion numbers~\cite{Callan:1983tm},
\begin{equation}
 e_R\, +\, M \,\to\, M\,+\,\frac{1}{2}\left((\overline{u})_{L1}\, (\overline{u})_{L2}\,(\overline{d}\,)_{R3}\, e_L\right)\,.
\label{eq:callan-intro}
\end{equation}
The outgoing state in~\eqref{eq:callan-intro} is built from half-solitons of the bosonised $s$-wave theory and has half-integer fermion number in each flavour; as a whole it carries half the baryon number of an anti-proton, and it is the massless precursor of the monopole-catalysed anti-proton synthesis process of~\cite{Khoze:2024hlb}.
Such a state does not belong to the Fock space in which the incoming fermions are defined, and its meaning has been debated since it was found.

We will also consider single-fermion-initiated processes with quarks in the initial state, such as
\begin{equation}
u_L^{\,1}+M\to M+\frac{1}{2}(u_R^{1}\,(\overline{u})_{L2}\,(\overline{d}\,)_{R3}\,(\overline{e})_{R})\,,\quad
u_L^{\,2}+M\to M+\frac{1}{2}(u_R^{2}\,(\overline{u})_{L1}\,(\overline{d}\,)_{R3}\,(\overline{e})_{R})\,.
\label{eq:1q}
\end{equation}
These can be viewed as elementary constituent processes behind~\eqref{eq:m-cat-p}. A combination of these two single-quark-initiated $j_0$-wave processes gives the more familiar two-quark-initiated monopole-catalysed 
proton decay process~\eqref{eq:m-cat-p}, allowing for annihilation of quark--anti-quark pairs $u_R^{1} (\overline{u})_{L1}$, $u_R^{2} (\overline{u})_{L2}$ into outgoing radiation.

Callan proposed a statistical interpretation~\cite{Callan:1983tm} of single-fermion processes~\eqref{eq:callan-intro}--\eqref{eq:1q}: the half-integer fermion numbers are expectation values in a superposition of ordinary Fock states, $|{\rm out}\rangle = \tfrac{1}{\sqrt2}\,|e_L\rangle + \tfrac{1}{\sqrt2}\,| (\overline{u})_{L1}\, (\overline{u})_{L2}\,(\overline{d}\,)_{R3}\rangle$, so that the $\Delta B=0$ and $\Delta B=-1$ channels each carry one half of the total cross-section. Once fermion masses are switched on, the vacuum left behind by the half-solitons sits midway between two minima of the sine-Gordon potential and must roll to one of them, so that at $r\gtrsim1/m$ the final state sorts itself into integer-fermion-number states. This reading was adopted in~\cite{Sen:1983yq,Sen:1984qe,Dawson:1983cm,Rubakov:1988aq,Preskill:1984gd}, and the fractional states came to be called semitons.

In the massless theory, however,  no state with integer fermion numbers carries the quantum numbers required by angular momentum, electric charge and the flavour symmetry~\cite{Kitano:2021pwt}, so a simple superposition of ordinary perturbative fermion Fock states cannot be the answer there. 

The strictly massless problem has recently been settled in~\cite{vanBeest:2023dbu}, using the conformal boundary states of~\cite{Smith:2019jnh} and the language of topological symmetry lines, and a similar conclusion has been reached  within the fermion--rotor model of~\cite{Polchinski:1984uw} in~\cite{Hamada:2022eiv,Loladze:2024ayk,Loladze:2025jsq,Geng:2026xob}.
The outgoing radiation lives in a different -- twisted -- sector of the Hilbert space~\cite{Hamada:2022eiv,vanBeest:2023dbu,vanBeest:2023mbs,Loladze:2025jsq}: in the formulation of~\cite{vanBeest:2023dbu} an ordinary outgoing fermion is attached to a topological line that connects it to the monopole,
\begin{equation}
 e_R\,+\, M\,\to\, M\,+\,Te_L\,.
 \label{eq:TprocN}
\end{equation}
Here $T=\frac{1}{2}\left((\overline{u})_{L1}\, (\overline{u})_{L2}\,(\overline{d}\,)_{R3} \,(\overline{e})_{R}\right)$ denotes the twist operator at the $e_L$ end of the topological line, whose other end is at $r=0$, on the monopole. The combination $Te_L$ on the right-hand side is meant to represent a single operator, not a product of two states, and its fractional charges are precisely those of Callan's half-soliton configuration~\eqref{eq:callan-intro}.
 In the massless limit \eqref{eq:callan-intro} and~\eqref{eq:TprocN} are therefore the same statement in two languages, related by bosonisation: on one side a fermion connected to the monopole by a topological string, on the other the semiton configuration of the sine-Gordon fields.

Scattering amplitudes for the process~\eqref{eq:callan-intro}, \eqref{eq:TprocN} were constructed 
in~\cite{Khoze:2023kiu} using the pairwise-helicity spinors~\cite{Csaki:2020inw} with fractional powers of spinor products, whose branch cuts are the image of the topological line.
Subsequently, the authors of~\cite{Cordova:2026dtw} computed the boundary correlator for the process~\eqref{eq:TprocN} and argued that the twisted channel saturates $s$-wave unitarity with the same strength as the ordinary helicity flip produced by the vector-like boundary condition, i.e.\ the lowest-partial-wave process of Kazama, Yang and Goldhaber (KYG) for a fermion scattering on a monopole in QED~\cite{Kazama:1976fm}.
Alternative and complementary studies include~\cite{Csaki:2021ozp,Hook:2024vhf,Csaki:2024ajo,Koundouros:2026rnb,Tachikawa:2026cxd,Hamada:2026jgj} and applications to the Standard Model are treated in Refs.~\cite{vanBeest:2023mbs,Khoze:2024hlb,Alonso:2025rkk,Alonso:2026vnr,Craig:2026rdf}.

\medskip
The massless theory thus has a consistent and unitary answer, but it is not the answer an experiment sees. In the Standard Model the fermions are massive and the quarks confine, and at $r\gtrsim1/m$ the topological line ceases to be topological: the twisted state must resolve itself into physical final states with integer fermion numbers. For an incoming electron there are only two options, a helicity flip or conversion into an anti-nucleon,
\begin{equation}
e_R \,+\, M \,\to\, M\,+\,e_L\,+\,X\,,\qquad\qquad
e_R \,+\, M \,\to\, M\,+\,\overline{p}_L\,+\,X\,,
\label{eq:two-intro}
\end{equation}
the second being the anti-proton synthesis process noted in~\cite{Khoze:2024hlb} and explored in~\cite{Candela:2025gwp} as the signature of relativistic monopoles in large neutrino detectors. What fraction of the unitarity-saturating rate goes into each channel, and with what kinematics, is the question this paper addresses.

\medskip
Our strategy is to use both descriptions, \eqref{eq:callan-intro} and \eqref{eq:TprocN}, of the same $s$-wave theory process, each addressing a complementary question, and to match them at $r\ll1/m$ where both apply.\footnote{To keep our discussion simple, we use $m$ here as a generic name for relevant mass scales of the theory -- confinement, quark condensate, fermion masses or the Higgs vev.} The boundary-CFT construction fixes the normalisation of the twisted channel and identifies the semitons in the final state as a single endpoint operator. The bosonised theory answers the one question the massless theory cannot, what the mass terms in the theory do to the final state.
The decay of the topological string, i.e.\ the twist line, is caused by the presence of mass scales in the theory, but it is an $O(1)$ effect, not captured at any finite order in $m$. The small parameter $m/E$ controls only the light-like propagation of the hard outgoing front, and hence the accuracy of the symmetry that fixes the branching. Beyond the twisted-sector description, the key input is that a Dirac mass term is odd under the twist line.
We will establish the following results, for one generation of fermions on $M$.\footnote{The extension to three generations is left for future work, see section~\ref{Sec:concl}.}
\begin{enumerate}
\item The overall $j_0$-wave cross-section is given by the KYG expression, $\sigma_{j_0}=\pi/p^2$, as anticipated in~\cite{Khoze:2024hlb} from unitarity saturation; it follows from the unit-modulus correlator of~\cite{Cordova:2026dtw} (section~\ref{sec:norm}).
\item The energy of the outgoing state in~\eqref{eq:callan-intro}--\eqref{eq:1q} is shared among the four $s$-wave flavours with a Dirichlet distribution. The mean shares are equal, but in a typical event a single sector carries more than half of the energy. This sets the kinematics and thresholds of the physical processes~\eqref{eq:two-intro} (section~\ref{sec:dirichlet}).
\item At $r\ll1/m$ the outgoing state is a single twisted state, not a superposition of two Fock states. It decays at $r\gtrsim1/m$, and conservation laws leave exactly the two channels~\eqref{eq:two-intro}: the line retracts into the monopole, which can absorb its half-unit of charge with either sign, $\pm\frac12(1,1,1,1)$, and the two outcomes differ by $\Delta B=\Delta L=1$ (section~\ref{sec:channels}).

\item The branching ratios for the two processes~\eqref{eq:two-intro} are fixed in the two limits. For $E\gg m_p$ the two channels are equally likely, by the charge-conjugation symmetry of the decaying string, which is insensitive to the hierarchy $m_e\ll\Lqcd$; these are Callan's weights of $1/2$, and they are specific to the four-flavour theory of the minimal monopole with one generation, in which the twist operator carries charge $1/2$ in every flavour. Below the proton-mass threshold, $E<m_p$, the incoming electron can only flip its helicity, and for $E\ll\Lqcd$ the quark sectors decouple altogether (sections~\ref{sec:high} and~\ref{sec:low}). The equal weights are a quantum effect. They rely on the fluctuations of the string dominating a classical bias of order $m/E$, which is estimated but not proven (section~\ref{sec:quantum}).
\item Callan's superposition of the two integer states does not describe the state at $r\ll1/m$, where the flavour numbers are sharp and half-integer. It is recovered for the asymptotic state at $r\gtrsim1/m$, with the weights replaced by the energy-dependent branching ratios and with each component dressed by neutral radiation. Since the two channels differ in baryon and lepton number, their relative phase is not observable and the superposition is equivalent to a statistical mixture (section~\ref{Sec:concl}).
\item For an incoming electron at high energies, approximately half of all events contain an energetic helicity-flipped electron in the final state, carrying more than $10\%$ of the incoming energy. In the remaining events no energetic electrons are present. This holds equally for the helicity-flip channel and for the anti-nucleon synthesis channel. In particular, about half of the anti-nucleon events also contain an energetic electron, which is a new signature. Relative to~\cite{Candela:2025gwp}, the rate of anti-nucleon synthesis is lower near the threshold (sections~\ref{sec:hard} and~\ref{Sec:lepton}).
\item For a slow proton the catalysed decay has a probability close to one, whatever the branching ratio of a single quark (section~\ref{Sec:proton}).
\end{enumerate}
The interpolating form of the branching ratio between its two limits remains a model, now constrained by the energy distribution. What is new relative to~\cite{Khoze:2024hlb} is that the overall factor, the channels and the limiting weights follow from the structure of the theory, and that the two-body kinematics assumed there acquires a broad event-by-event spread.

\medskip
The paper is organised as follows. Section~\ref{Sec:setup} collects what is needed from~\cite{Khoze:2024hlb,Cordova:2026dtw} and fixes the bosonised notation. Section~\ref{Sec:massless} contains the exact results of the massless theory, section~\ref{Sec:mass} describes what the masses do to the twisted state, and section~\ref{Sec:br} derives the branching ratios. Section~\ref{Sec:phys} spells out the consequences for electron- and proton-initiated processes, and section~\ref{Sec:concl} provides the summary. Appendix~\ref{app:reflection} repeats the symmetry argument of section~\ref{sec:high} in bosonised form, appendix~\ref{app:hard} gives the details of the computation behind section~\ref{sec:hard}, and appendix~\ref{sec:classical} describes the simulations of the classical bosonised model that are summarised in section~\ref{sec:quantum}.

\medskip
\medskip
\noindent{\bf Note added.} In Ref.~\cite{Tong:2026apr}, which was posted on the arXiv on the same day as the first version of this paper, the outgoing twisted state of the massless theory was analysed from the point of view of a four-dimensional observer. In the regime where the two papers overlap, their main conclusions appear to be in agreement. 

\section{Fermions in the $j_0$-wave of the minimal monopole}
\label{Sec:setup}

\subsection{Fermion content and the KYG cross-section}
\label{sec:kyg}

In the $j_0$-wave scattering on the minimal monopole $M$, a single generation of Standard Model fermions reduces to four Dirac fermions~\cite{Rubakov:1988aq,vanBeest:2023mbs,Khoze:2024hlb}, listed in Table~\ref{Tab:M1ferm}. Each has a left- or right-handed component that moves towards the core (incoming, $\psi_i$) and the opposite-helicity component that moves away from it (outgoing, $\tilde\psi_i$). 
This assignment is fixed by the sign of the pairwise helicity $q_J$ -- the Dirac-quantised product of the electric charge of the fermion and the magnetic charge of the monopole. For every flavour $|q_J|=1/2$, so $j_0=|q_J|-\frac{1}{2}=0$ and the $s$-wave is literally a two-dimensional theory of four Dirac fermions on the half-line $r>0$.

\begin{table}[ht]
\begin{equation} \nonumber
\begin{array}{c|cc|ccc}
\,\,i\,\, &\quad \psi_i\,\, \text{incoming}\quad & \quad\tilde\psi_i\,\, \text{outgoing}\quad &\quad q_J\quad &\quad Q_{el}\quad &\quad B-L\quad
\\\hline
1 &\,\, e_R & e_L & -\frac{1}{2} & -1 & -1
\\
2 &\,\, u_L^{\,1} & u_R^{\,1} & +\frac{1}{2} & +\frac{2}{3} & +\frac{1}{3}
\\
3 &\,\, u_L^{\,2} & u_R^{\,2} & +\frac{1}{2} & +\frac{2}{3} & +\frac{1}{3}
\\
4 &\,\, d_R^{\,3} & d_L^{\,3} & -\frac{1}{2} & -\frac{1}{3} & +\frac{1}{3}
\end{array}
\end{equation}
\caption{The four Dirac fermions $(\psi_i,\tilde\psi_i)_{i=1,\dots,4}$ that interact with the minimal $SU(5)_{\rm GUT}$ monopole $M$ in the $j_0$-wave (Table~5 of~\cite{Khoze:2024hlb}). Each consists of an incoming Weyl fermion $\psi_i$ and an outgoing Weyl fermion $\tilde\psi_i$ of the opposite helicity. Upper indices are $SU(3)_c$ colours of quarks. A state of the $s$-wave theory is labelled by its flavour vector $\lambda\in\Z^4$, the fermion number in each of the four Dirac fields. A single $\psi_i$ or $\tilde\psi_i$ has $\lambda=\hat e_i$, the $i$-th unit vector.}
\label{Tab:M1ferm}
\end{table}

The lowest partial wave, $j=j_0$, is where the helicity-flip and $(B+L)$-violating processes catalysed by the monopole occur. All higher partial waves, $j>j_0$, scatter elastically, without reaching the core and without helicity flip or $(B+L)$ violation.

The $j_0$-wave cross-section was computed by KYG~\cite{Kazama:1976fm} for a single Dirac fermion scattering on a monopole in QED, where all the core can do is flip the helicity: $d\sigma_{j_0}/d\Omega=(q_J^2/p^2)\left(\sin\theta/2\right)^{4|q_J|-2}$. For the minimal monopole $|q_J|=1/2$, and
\begin{equation}
\frac{d\sigma_{j_0}}{d\Omega}\,=\, \frac{1}{4p^2}\,,\qquad \sigma_{j_0}\,=\,\frac{\pi}{p^2}\,,
\label{eq:sigKYG}
\end{equation}
in the frame in which the monopole is at rest, $p$ being the momentum of the incoming fermion. The cross-section is isotropic and, which is the key point, it saturates the unitarity bound on an inelastic process in the lowest partial wave, $\sigma_{\rm inel}\leq\pi(2j_0+1)/p^2=\pi/p^2$: every fermion that reaches the core returns in the new channel. The Callan--Rubakov processes of the full theory proceed in the same partial wave and at the same unitarity limit~\cite{Rubakov:1982fp,Callan:1982ac,Rubakov:1988aq}.

The proposal of~\cite{Khoze:2024hlb} for the Standard Model, with its mass scales, starts from Callan's statistical interpretation: each incoming fermion has two physical processes into asymptotic final states at $r\gg1/\Lqcd$, the helicity flip and its $(B+L)$-violating partner, and the cross-section of each is the unitarity-saturating expression~\eqref{eq:sigKYG} times a branching ratio. For an incoming electron,
\begin{equation}
\begin{aligned}
e_R + M &\,\to\, M+e_L\,+\,X\,: &\quad \sigma&\,=\,\sigma_{j_0}\,\br_{e}\,, &\quad \br_{e}&\,=\,1-\br_{\bar p}\,,\\[2pt]
e_R + M &\,\to\, M+\overline{p}_L\,+\,X\,: &\quad \sigma&\,=\,\sigma_{j_0}\,\br_{\bar p}\,, &\quad \br_{\bar p}&\,=\,\frac{1}{2}\,\frac{k_{\bar p}}{p}\,,
\end{aligned}
\label{eq:flipsig-e}
\end{equation}
where $k_{\bar p}=(E^2-m_p^2)^{1/2}$ is the momentum of the outgoing anti-proton, so that $\br_{\bar p}$ is one half of the ratio of the two-body phase-space volumes: it vanishes at the threshold $E=m_p$ and tends to Callan's $1/2$ at high energy. 

The second example is a non-relativistic proton, with velocity $\beta<0.2$. Its wavelength, $1/(m_p\beta)$, then exceeds the proton radius, so the proton scatters on the monopole as a single object in the $j_0$-wave, and not parton by parton. The momentum in~\eqref{eq:sigKYG} is that of the proton, $p=m_p\beta$, so that $\sigma_{j_0}=\pi/(m_p^2\beta^2)$. The two channels are the helicity flip of the proton and its catalysed decay into a positron,
\begin{equation}
\begin{aligned}
p_L + M &\,\to\, M+p_R\,+\,X\,: &\quad \sigma&\,=\,\sigma_{j_0}\,\br_{p}\,, &\quad \br_{p}&\,=\,\frac{k_{p}}{k_{\bar e}+k_{p}}\,,\\[2pt]
p_L + M &\,\to\, M+\overline{e}_R\,+\,X\,: &\quad \sigma&\,=\,\sigma_{j_0}\,\br_{\bar e}\,, &\quad \br_{\bar e}&\,=\,\frac{k_{\bar e}}{k_{\bar e}+k_{p}}\,,
\end{aligned}
\label{eq:flipsig-p}
\end{equation}
where $k_p$ and $k_{\bar e}$ are the momenta of the outgoing proton and positron, and the branching ratios are again taken proportional to the phase space of the final state. The outgoing proton is as slow as the incoming one, $k_p=m_p\beta$, while the positron carries away a large part of the proton's rest energy, $k_{\bar e}\sim m_p$. Hence $k_{\bar e}\gg k_p$ and the catalysed decay dominates, $\br_{\bar e}\approx1$.

Equations~\eqref{eq:flipsig-e}--\eqref{eq:flipsig-p} are a conjecture with three ingredients: that the two processes in each of~\eqref{eq:flipsig-e} and~\eqref{eq:flipsig-p} are the relevant channels, that the overall factor is the KYG cross-section, and that the branching ratios can be reasonably estimated from kinematics alone. The second is well motivated, since unitarity must be saturated in the lowest partial wave and this is precisely what~\eqref{eq:sigKYG} does. The first and the third refine Callan's statistical rule, replacing the equal weights of $1/2$ by energy-dependent branching ratios. In the present paper the overall factor will be derived (section~\ref{sec:norm}), the two channels will be confirmed (section~\ref{sec:channels}), and the branching ratios and the kinematics behind them will be addressed in detail (sections~\ref{Sec:br}--\ref{Sec:phys}).

\subsection{The boundary condition and the twisted state}
\label{sec:twisted}

The derivations below use Callan's bosonised form of the $s$-wave theory~\cite{Callan:1982au,Callan:1983tm}. 
For each flavour $j$ the fields $\psi_j$ and $\tilde\psi_j$ of Table~\ref{Tab:M1ferm} are the two chiral components of a single two-dimensional Dirac fermion on the half-line $r>0$, moving towards and away from the core. In the massless theory they decouple and are functions of $t+r$ and $t-r$ respectively. By the Coleman correspondence~\cite{Coleman:1974bu}, chiral fermions are identified with the chiral vertex operators of free scalars
in $(1+1)$ dimensions,
\begin{equation}
\psi_j\,\sim\,:\!e^{\,i\phi_j^{\rm in}}\!:\,,\qquad
\tilde\psi_j\,\sim\,:\!e^{-i\phi_j^{\rm out}}\!:\,.
\label{eq:chiral}
\end{equation}
The vertex operators have conformal weight $h=\frac{1}{2}$, the scaling dimension of a free fermion.
In the massless limit, the scalar theory is just the kinetic term, the general solution splits into chiral halves, and the fermion-number current of flavour $j$ (the Noether current of the flavour symmetry $U(1)_j$ under which $\psi_j$ and $\tilde\psi_j$ both have charge one) is topological,
\begin{equation}
{\cal L}\,=\,\sum_j \frac{1}{8\pi}\,(\partial\phi_j)^2\,,\qquad
\phi_j\,=\,\phi_j^{\rm in}(t+r)\,+\,\phi_j^{\rm out}(t-r)\,,\qquad
j^\mu_j\,=\,\frac{1}{2\pi}\,\epsilon^{\mu\nu}\partial_\nu\phi_j\,.
\label{eq:bosonisation}
\end{equation}
Introducing the Dirac mass term for fermions corresponds to adding to the Hamiltonian density the operator,
\begin{equation}
m_j\big(\psi_j\tilde\psi_j^\dagger+{\rm h.c.}\big)\;\to\;-\frac{m_j\Lambda}{\pi}\,\cos\phi_j\,,
\label{eq:mass}
\end{equation}
of dimension one, with $\Lambda$ the normal-ordering scale; the coefficient is scheme dependent and only its sign and the period of the cosine matter below. Thus the dual bosonised description of massive Dirac fermions is the sine-Gordon model at Coleman's coupling $\beta^2=4\pi$, the free-fermion point. The sine-Gordon soliton is the fermion itself and there are no breathers.
The fermion number in an interval $r_1<r<r_2$ is the winding of $\phi_j$ across it in units of $2\pi$, $N_j[r_1,r_2]=\int_{r_1}^{r_2}dr\,j^0_j=\frac{1}{2\pi}\big[\phi_j(r_2)-\phi_j(r_1)\big]$: a fermion is a $2\pi$ kink and Callan's half-soliton a $\pi$ kink.

 The chiral halves $\phi_j^{\rm in}$, $\phi_j^{\rm out}$ are related to the fermions $\psi_j$, $\tilde\psi_j$ by the dictionary~\eqref{eq:chiral}.
 They carry the fermion-number densities
\begin{equation}
\rho^{\rm in}_j\,=\,\frac{1}{2\pi}\,\frac{\partial\phi^{\rm in}_j}{\partial(t+r)}\,,\qquad
\rho^{\rm out}_j\,=\,-\frac{1}{2\pi}\,\frac{\partial\phi^{\rm out}_j}{\partial(t-r)}\,,
\label{eq:rho}
\end{equation}
whose sum is the density $j^0_j=\frac{1}{2\pi}\partial_r\phi_j$ of~\eqref{eq:bosonisation}.
 In general, a chiral vertex operator built from the $N_f=4$ outgoing fields of Table~\ref{Tab:M1ferm}
 is
 \begin{equation}
{\cal O}_\lambda\,=\,e^{-i\lambda\cdot\phi^{\rm out}}\,,\qquad \lambda\in\mathbb{R}^4\,,\qquad h\,=\,\tfrac{1}{2}|\lambda|^2\,.
\label{eq:vertex}
\end{equation}
It carries flavour vector $\lambda$; it is a local operator of the fermion theory for $\lambda\in\Z^4$, and for the semiton states the entries of $\lambda$ are half-integer. 

\medskip

The $s$-wave bulk is the sum of $N_f$ free fields ($N_f=4$ for one generation on the minimal monopole $M$), and the flavours talk to each other only through the boundary at $r=0$. The massless core preserves the flavour symmetry, in particular its $U(1)^{N_f}$ subgroup generated by
\begin{equation}
N_j\,=\, \int dr\,j^0_j \,=\, \frac{1}{2\pi}\int dr\,\partial_r\phi_j\,, \qquad j=1,\ldots,N_f\,,
\end{equation}
and the Callan--Rubakov boundary condition is linear in this language~\cite{vanBeest:2023dbu}: it glues the outgoing fermion-number densities to the incoming ones by an orthogonal matrix,
\begin{equation}
\rho^{\rm out}_j\,=\,\sum_{k=1}^{N_f} R_{jk}\,\rho^{\rm in}_k\quad\text{at }r=0\,,\qquad R\,=\,\one-\frac{2}{N_f}\,J\,=\,\one-\frac{1}{2}J\quad(N_f=4)\,,
\label{eq:R}
\end{equation}
where $J$ is the matrix of ones. 

For a larger number of flavours the form $\one-\frac{2}{N_f}J$ holds in the basis of the $SU(2)_M$ doublets,\footnote{Here $SU(2)_M$ is the $SU(2)$ subgroup of the gauge group in which the monopole is embedded. For the minimal monopole of $SU(5)_{\rm GUT}$ it mixes the third colour index of the fundamental representation with one weak index, which is why the down quark appears in Table~\ref{Tab:M1ferm} with colour 3 and the up quarks, which belong to the antisymmetric representation, with colours 1 and 2. Written as left-handed Weyl fields, the fermions of Table~\ref{Tab:M1ferm} form four doublets of $SU(2)_M$:
$\big((\overline{e})_{L}\,,\,d^{\,3}_L\big)$, $\big((\overline{d}\,)_{L3}\,,\,e_L\big)$, $\big(u^{\,1}_L\,,\,(\overline{u})_{L2}\big)$ and $\big(u^{\,2}_L\,,\,(\overline{u})_{L1}\big)$,
with the first component incoming and the second outgoing. A doublet thus pairs an incoming fermion with an outgoing fermion of a different Dirac flavour, whereas a row of Table~\ref{Tab:M1ferm} pairs the two components of the same flavour.} each made of an incoming and an outgoing Weyl fermion~\cite{vanBeest:2023dbu}, and not in the basis of the Dirac flavours of Table~\ref{Tab:M1ferm}, in which the mass terms are diagonal. 
The two forms differ only in which outgoing fermion carries the weight $1-\frac{2}{N_f}$, the partner of the incoming fermion in its doublet or its helicity-flipped state. For $N_f=4$ this weight is equal to the weight $\frac{2}{N_f}$ of the others, so the two forms coincide. This is the only case used in this paper.

Since $\partial_t\phi_j=2\pi(\rho^{\rm in}_j-\rho^{\rm out}_j)$ and $\partial_r\phi_j=2\pi(\rho^{\rm in}_j+\rho^{\rm out}_j)$ at $r=0$, a combination of the fields on which $R$ has eigenvalue $+1$ is Dirichlet ($\partial_t\phi=0$, no current into the core, charge reflected) and one on which it has eigenvalue $-1$ is Neumann ($\partial_r\phi=0$, charge absorbed by the core and re-emitted with the opposite sign).
$R=\one$ is the charge-conserving vector-like boundary of KYG, Dirichlet for every flavour. The Callan--Rubakov matrix~\eqref{eq:R} is Dirichlet on the differences $\phi_i-\phi_j$, so all four fields share one boundary value $\phi_b(t)$ and the differences of the flavour numbers are conserved, and Neumann on the sum $\sum_j\phi_j$, so the total fermion number is absorbed: it conserves the vector $U(1)$ (electric charge) and the non-anomalous flavour combinations ($B-L$), and violates the anomalous chirality count ($B+L$) uniformly in all flavours.

Applied to the flavour vector, $\lambda\to R\lambda$. For a single incoming $\psi_i$ the flavour vector is $\hat e_i$, the unit vector with $1$ in the $i$-th entry and $0$ in the others, and $R\hat e_i\,=\,\hat e_i-\frac{1}{2}(1,1,1,1)$: the fermion comes out as half a unit of $\tilde\psi_i$ together with half a unit of anti-fermion in each of the other three fields,
\begin{subequations}
\label{eq:callan}
\begin{align}
 u_L^{\,1}\, +\, M \,&\to\, M\,+\,\frac{1}{2}\left(u_R^{1}\, (\overline{u})_{L2}\,(\overline{d}\,)_{R3} \,(\overline{e})_{R}\right)\,,
\label{eq:callan-u}\\
d_R^{\,3}\, +\, M \,&\to\, M\,+\,\frac{1}{2}\left(d_L^{\,3} \,(\overline{u})_{L1}\, (\overline{u})_{L2}\,(\overline{e})_{R}\right)\,,
\label{eq:callan-d}\\
 e_R\, +\, M \,&\to\, M\,+\,\frac{1}{2}\left(e_L\,(\overline{u})_{L1}\, (\overline{u})_{L2}\,(\overline{d}\,)_{R3} \right)\,,
\label{eq:callan-e}
\end{align}
\end{subequations}
and $u_L^{\,2}$ as $u_L^{\,1}$ with $1\leftrightarrow2$. These are Callan's half-soliton, or semiton, states~\cite{Callan:1983tm} already 
quoted in the Introduction in~\eqref{eq:callan-intro}--\eqref{eq:1q}. 

The outgoing state is the chiral vertex operator ${\cal O}_i\equiv{\cal O}_{R\hat e_i}$ of~\eqref{eq:vertex}, labelled by the incoming flavour $i$ and carrying the outgoing charges $R\hat e_i$.
 The charges are half-integer, so it is not a local operator of the untwisted theory, and its weight is $\frac{1}{2}|R\hat e_i|^2=\frac{1}{2}$, that of a single fermion. 
 
 What it is instead is made precise in~\cite{vanBeest:2023dbu,Cordova:2026dtw}. Each conserved charge can be exponentiated into a topological line operator, $\exp(i\sum_j\theta_jN_j)$ inserted along a curve, which can be moved freely except across charged operators, and an operator may sit at the end of such a line. 
 A fermion carried around that endpoint picks up the phase $e^{i\theta_j}$, so the endpoint operator is not mutually local with the fermions and belongs to a twisted Hilbert space rather than the ordinary one. For ${\cal O}_i$ the phases are $e^{2\pi i(R\hat e_i)_j}=-1$ for every $j$, so the line is $D=\exp(i\pi\sum_jN^{\rm out}_j)$, the $(-1)^F$ of the outgoing fermions, running from the operator back to the boundary, and the semiton lives in the $D$-twisted sector. The twisted sectors form $\Z^{N_f}/\Gamma$ with $\Gamma=\{\lambda:R\lambda\in\Z^{N_f}\}=\{\lambda:\sum_j\lambda_j\in\frac{N_f}{2}\Z\}$, i.e.\ $\Z_{N_f/2}$. 
 For $M$ this is $\Z_2$: a state is in the twisted sector when the total incoming fermion number is odd, and $D$ is the $\Z_2$ generator. For a single incoming fermion the outgoing charge in each of the other flavours is $-2/N_f$, and we refer to its magnitude $2/N_f$ as the twist charge. It is $\frac{1}{2}$ for $M$. In the fermion--rotor description of the core~\cite{Polchinski:1984uw} the same line is the rotor vertex $e^{i\alpha(t-r)}$ that dresses the outgoing fermion, whose commutator with the fermion field reduces in the infrared to the action of $D$~\cite{Loladze:2024ayk,Loladze:2025jsq,Geng:2026xob}.
In~\cite{Khoze:2024hlb}, following~\cite{vanBeest:2023dbu,vanBeest:2023mbs}, the same state was written as ({\it cf.}~\eqref{eq:TprocN}),
\begin{equation}
 \psi_i\,+\, M\,\to\, M\,+\,T\,\tilde\psi_i\,,\qquad
 T \,=\,  \frac{1}{2}\,\prod_{j=1}^{4} \tilde\psi^{\,\dagger}_j\,=\,
 \frac{1}{2}\left((\overline{u})_{L1}\, (\overline{u})_{L2}\,(\overline{d}\,)_{R3} \,(\overline{e})_{R}\right)\,.
 \label{eq:Tproc1}
\end{equation}
Here $T$ is the twist operator at the end of the line. It carries the flavour vector $-\frac{1}{2}(1,1,1,1)$, the quantum numbers of half an anti-proton and half a positron, so that~\eqref{eq:Tproc1} is the identity $R\hat e_i=\hat e_i-\frac{1}{2}(1,1,1,1)$ written in terms of operators. The combination ${\cal O}_i=T\,\tilde\psi_i$ is a single vertex operator, not a product of two states.
Scattering amplitudes for processes~\eqref{eq:callan} were constructed in~\cite{Khoze:2023kiu}; they involve spinor products raised to the power $1/2$, the branch cut of the square root being the line $D$ on the $(r,t)$ half-plane.

\section{Exact results of the massless theory}
\label{Sec:massless}

\subsection{The overall cross-section}
\label{sec:norm}

The massless $j_0$-wave process is the one-to-one transition of~\eqref{eq:Tproc1},
\begin{equation}
\psi_i(E)\,+\,M\,\to\,M\,+\,{\cal O}_i(E)\,.
\label{eq:process}
\end{equation}
Its two-dimensional avatar is a boundary two-point function. We use Euclidean coordinates on the half-plane, $z=\tau+ir$ with $\tau=it$ and $r\geq0$, so that the core $r=0$ is the real axis. The incoming fermion is holomorphic, a function of $z$, and the outgoing operators are antiholomorphic, functions of $\bar z$. The two-point function of the incoming fermion at $z$ with the conjugate of the outgoing twisted operator at $w$, computed in~\cite{Cordova:2026dtw} from the twisted boundary state, is
\begin{equation}
\big\langle\psi_i(z)\,{\cal O}^\dagger_i(\bar w)\big\rangle_B\,=\,\frac{1}{|z-\bar w|}\,,
\label{eq:twopt}
\end{equation}
with a coefficient of unit modulus, the same as for the vector-like boundary that produces the KYG helicity flip $\psi_i\to\tilde\psi_i$. The dependence on $z-\bar w$ alone is the image structure of a boundary correlator, its pole at $t'-t=r+r'$, with $(t',r')$ the position of the outgoing operator, being the light ray reflected off the core. The power $2h=1$ follows from the weights $h=\frac12$ of both operators.

Ref.~\cite{Cordova:2026dtw} relates such correlators to four-dimensional scattering through the reduced $S$-matrix element $S(\hat p,\hat q)$. It is defined by stripping the energy delta function from the one-to-one $S$-matrix element between gauge-covariant momentum states, ${\cal M}(\vec p,\vec q)=4\pi^2\delta(p-q)\,p^{-2}\,S(\hat p,\hat q)$, and unitarity reads $\int\frac{d^2\hat q}{(2\pi)^2}S^\dagger S=\delta^{(2)}(\hat p'-\hat p)$. In these conventions the free $S$-matrix is $S_{\rm free}=2\pi\delta^{(2)}(\hat q-\hat p)$, whose $\ell=0$ component is $1/2$. The bulk Dirac equation alone gives an $S$-matrix that violates unitarity by exactly the $s$-wave projector. For the vector-like boundary the authors of~\cite{Cordova:2026dtw} derive that the boundary condition adds $S_{\psi_i\to\tilde\psi_i}=1/2$. This restores unitarity, since $\int\frac{d^2\hat q}{(2\pi)^2}|S|^2=\frac{1}{4}\cdot\frac{4\pi}{4\pi^2}=\frac{1}{4\pi}$ is the missing projector. For the Callan--Rubakov boundary the outgoing state has no four-dimensional field description, and they propose, by the same identification, that the unit-modulus correlator~\eqref{eq:twopt} gives
\begin{equation}
S_{\psi_i\to{\cal O}_i}\,=\,\frac{1}{2}\,,
\label{eq:Shalf}
\end{equation}
so that the entire $s$-wave goes into the twisted state with probability one and unitarity is again restored.

The result~\eqref{eq:Shalf} can in fact be made independent of the conventions of the boundary-correlator/$S$-matrix dictionary. Three facts fix it. First, the bulk Dirac equation in the monopole background gives no $j_0$-wave reflection at all, so whatever returns in the $s$-wave is emitted by the boundary. Second, the boundary state is determined by the symmetries it must preserve~\cite{Smith:2019jnh,vanBeest:2023dbu}, up to phases that do not affect probabilities, so there is one boundary and not a family with a free parameter. Third, for a single incoming fermion the only operator that this boundary can emit with the right charges is ${\cal O}_i$, so the twisted channel is the unique open one. Conservation of probability in the $j_0$-wave then forces $|S_{\psi_i\to{\cal O}_i}|^2=1/4$, whatever the normalisation of the correlators. The role of~\eqref{eq:twopt} is to confirm that the twisted operator is emitted with the full strength and not, say, partly reflected into a different sector. The same reasoning applied to the vector-like boundary reproduces the KYG result, which calibrates the dictionary.

Nothing in the dynamics of the monopole core obstructs this at low momentum. In the fermion--rotor description the charge deposited on the core is carried by the rotor. An isolated rotor has an excitation gap of order $1/I$, with $I$ its moment of inertia, but the rotor vertex of the coupled fermion--rotor system has spectral weight at arbitrarily small energy~\cite{Geng:2026xob}. There is therefore no threshold set by the dyon tower, and the twisted channel is open for arbitrarily small $p$.

The cross-section follows from the standard partial-wave relation. Writing $S=\sum_\ell\frac{2\ell+1}{2}s_\ell P_\ell$ and comparing with $f=\sum_\ell(2\ell+1)\frac{s_\ell-1}{2ip}P_\ell$ gives $f=(S-S_{\rm free})/(ip)$. Since the twisted channel has no free part,
\begin{equation}
\frac{d\sigma_{j_0}}{d\Omega}\,=\,\frac{|S_{\psi_i\to{\cal O}_i}|^2}{p^2}\,=\,\frac{1}{4p^2}\,,\qquad \sigma_{j_0}\,=\,\frac{\pi}{p^2}\,,
\label{eq:sigma0}
\end{equation}
which is~\eqref{eq:sigKYG} at $|q_J|=1/2$. The twisted channel of $M$ thus has the KYG cross-section, and this is the overall factor anticipated in~\eqref{eq:flipsig-e}--\eqref{eq:flipsig-p}. The result is consistent with the amplitude of~\cite{Khoze:2023kiu}. Every pairwise spinor product there has modulus $2E$, so after normalising each by the energy of the corresponding particle the $s$-wave amplitude for $\psi_i\to{\cal O}_i$ has unit modulus.
The outgoing state is a single state of the $s$-wave theory, a function of $t-r$ alone, so there is no multi-body phase space in the cross-section.

Since the massless theory produces the single state~\eqref{eq:process} with probability one, every physical cross-section is~\eqref{eq:sigma0} times the probability that this state, once the masses act on it, ends as a given physical final state ${\cal F}$,
\begin{equation}
\sigma(\psi_i+M\to M+{\cal F})\,=\,\sigma_{j_0}(p)\times\br_{\cal F}(E)\,,\qquad\sum_{\cal F}\br_{\cal F}=1\,.
\label{eq:factor}
\end{equation}
This is the form in which the cross-sections were written in~\cite{Khoze:2024hlb}. There it was an ansatz. Here it is the definition of $\br_{\cal F}$, and the rest of the paper is about these numbers.

\subsection{How the energy is shared among the flavours}
\label{sec:energy}

In the massless region the bulk $s$-wave theory is a sum of decoupled flavour sectors, the Lagrangian~\eqref{eq:bosonisation} being a sum over flavours, and the flavours talk to each other only through the boundary condition~\eqref{eq:R}. The Dirac mass terms~\eqref{eq:mass} do not change this, since each involves a single flavour. The strong interactions do, and at $r\gtrsim1/\Lqcd$ only the split between the electron sector and the three quark sectors taken together remains meaningful.

The energy in flavour $j$ is the sum of the energies of its incoming and outgoing halves, each quadratic in the corresponding fermion-number density,
\begin{equation}
H_j\,=\,\frac{1}{8\pi}\int dr\,\big[(\partial_t\phi_j)^2+(\partial_r\phi_j)^2\big]\,=\,\pi\int dr\,\big[(\rho^{\rm in}_j)^2+(\rho^{\rm out}_j)^2\big]\,,
\label{eq:Hj}
\end{equation}
with normal ordering understood. Each $H_j$ is separately conserved in the massless bulk. It also commutes with the line $D$, which is built from the $U(1)_j$ charges, so it is a well-defined operator in the twisted sector.

Consider an incoming fermion of energy $E$ in flavour 1, the electron. 
The boundary condition~\eqref{eq:R} gives $\rho_j^{\rm out}=\sum_kR_{jk}\,\rho_k^{\rm in}$, so the outgoing energy in flavour $j$ involves the products $R_{jk}R_{jl}\,\rho_k^{\rm in}\rho_l^{\rm in}$. The incoming flavours other than the first are in their vacuum, where the fermion-number density has zero expectation value and the normal-ordered energy vanishes. The only product that contributes to the expectation value is therefore the one with $k=l=1$, which is $R_{j1}^2$ times the energy of the incoming fermion. Hence
\begin{equation}
\frac{\langle E_j^{\rm out}\rangle}{E}\,=\,R_{j1}^2\,=\,\frac{1}{4}\quad\text{for every }j\,,\qquad \sum_jR_{j1}^2=1\,,
\label{eq:E4}
\end{equation}
where the sum rule, which expresses energy conservation, is the orthogonality of $R$. Charge goes as the first power of the same matrix element, $N_j^{\rm out}=R_{j1}=\pm\frac{1}{2}$, and energy as its square. On average the energy of the outgoing state is therefore equipartitioned: the electron sector and each of the three anti-quark sectors carry $E/4$, and the same holds for any incoming flavour. This is a statement about the mean energy per flavour, exact in the massless region.

Equation~\eqref{eq:E4} is consistent with the twist-operator form~\eqref{eq:Tproc1}, but it corrects a reading that one might attach to it, that ``$\tilde\psi_i$ carries the energy and $T$ carries none''. The line $D$ is a symmetry defect. It commutes with the stress tensor and indeed carries no energy. But the outgoing state is not the line. It is the operator at its endpoint, and that is a single object: $T\,\tilde\psi_i$ is a product of two vertex operators at the same point, not a two-body state. Conformal weights do not add in such a product. The weight of ${\cal O}_i$ is $\frac{1}{2}$ (section~\ref{sec:twisted}), while the weights of $\tilde\psi_i$ and $T$ taken separately add up to $\frac{1}{2}+\frac{1}{2}=1$. The endpoint operator therefore carries all of $E$, shared among the four sectors as in~\eqref{eq:E4}.

\subsection{The full distribution of the energy}
\label{sec:dirichlet}

The products that do not contribute to the mean in~\eqref{eq:E4} do contribute to the variance, and the fluctuations of the partition turn out to be as large as the means. The full distribution can be computed exactly.

We use box normalisation, in which the momenta $k$ of the modes of the boson fields are positive integers in units of $2\pi/L$, and the energy $E$ is an integer in the same units. In terms of canonically normalised oscillators, $[a_{j,k},a_{j',k'}^\dagger]=\delta_{jj'}\delta_{kk'}$, the incoming fields are
\begin{equation}
\phi_j^{\rm in}\,=\,\sum_{k\geq1}\frac{1}{\sqrt{k}}\,\Big(a_{j,k}\,e^{-ik(t+r)}+a_{j,k}^\dagger\,e^{ik(t+r)}\Big)\,,
\label{eq:modes}
\end{equation}
where $1/\sqrt{k}$ comes from the usual factor $1/\sqrt{2\omega_k}$ of a free massless field in one spatial dimension. The boundary condition~\eqref{eq:R} acts mode by mode. For every $k$ it rotates the oscillators $a_{j,k}$ of the four sectors by the orthogonal matrix $R$, a four-port beam splitter.

In bosonised form an incoming fermion of energy $E$ in flavour 1 is the component of the vertex operator $e^{i\phi_1^{\rm in}}$ with total momentum $E$, energy and momentum being equal for a massless chiral excitation. Acting on the vacuum, the vertex operator creates a coherent state,
\begin{equation}
:\!e^{\,i\phi_1^{\rm in}}\!:|0\rangle\,=\,\exp\Big(i\sum_{k\geq1}\frac{1}{\sqrt{k}}\,a_{1,k}^\dagger\,e^{ik(t+r)}\Big)|0\rangle\,,
\label{eq:coherent}
\end{equation}
in which the mode $k$ has an amplitude of modulus $1/\sqrt{k}$. Its component of energy $E$ is a superposition of the bosonic Fock states with occupation numbers $\{n_k\}$ that satisfy $\sum_kk\,n_k=E$, so that the energy is shared among quanta of momenta $k\leq E$, and the probabilities of these states are $\prod_kk^{-n_k}/n_k!$. To see that they sum to one for every $E$, we introduce a counting variable $\xi$ whose power records the energy, and sum over all occupation numbers without the constraint,
\begin{equation}
\sum_{\{n_k\}}\,\xi^{\,\sum_kk\,n_k}\,\prod_k\frac{k^{-n_k}}{n_k!}\,=\,\exp\Big(\sum_{k\geq1}\frac{\xi^k}{k}\Big)\,=\,\frac{1}{1-\xi}\,=\,\sum_{E\geq0}\xi^E\,.
\label{eq:cauchy}
\end{equation}
The states of energy $E$ are collected in the coefficient of $\xi^E$, which is one. The same probabilities describe the cycle lengths of a random permutation of $E$ objects (Cauchy's formula), with $n_k$ the number of cycles of length $k$, so the statistics of the cloud of quanta are known. The mean occupation number of the mode $k$ is the squared amplitude, $1/k$. 
The hardest quantum carries on average $62\%$ of the energy (the Golomb--Dickman constant), and the next one $21\%$.\footnote{The first number can be understood as follows. At most one quantum can have $k>E/2$, since two would exceed the total energy. The probability that such a quantum is present is therefore the mean number of quanta above $E/2$, $\sum_{k>E/2}1/k=\ln2\simeq0.69$, and its contribution to the mean share of the hardest quantum is $\sum_{k>E/2}(k/E)(1/k)=\frac{1}{2}$. So in $69\%$ of the states one quantum carries more than half of the energy, and these states alone contribute $0.50$ to the average. The remaining $31\%$ of the states, in which the hardest quantum is below $E/2$, contribute a further $0.12$, which gives the total of $0.62$.}
These are quanta of the bosonised field and not particles. The fermionic content of the outgoing state is discussed in section~\ref{sec:hard}.

The beam splitter sends each quantum, whole, to sector $j$ with probability $R_{j1}^2$, independently. With a counting variable $\xi_j$ for the energy of the sector $j$, the generating function of the sector energies is therefore $\prod_j(1-\xi_j)^{-R_{j1}^2}$. The coefficient of $\prod_j\xi_j^{E_j}$ is $\prod_j\Gamma(E_j+R_{j1}^2)/\big(\Gamma(R_{j1}^2)\,E_j!\big)$, which for large $E_j$ behaves as $\prod_jE_j^{\,R_{j1}^2-1}/\Gamma(R_{j1}^2)$, so that in the continuum limit the fractions $x_j=E_j/E$ are distributed on the simplex $\sum_jx_j=1$ with the Dirichlet density
\begin{equation}
P(x_1,\dots,x_4)\,=\,\frac{1}{\prod_j\Gamma(R_{j1}^2)}\,\prod_j x_j^{\,R_{j1}^2-1}\,=\,\frac{1}{\Gamma(\frac14)^4}\,\prod_{j=1}^4 x_j^{-3/4}\,,
\label{eq:dirichlet}
\end{equation}
i.e.\ Dirichlet$\big(\tfrac{1}{4},\tfrac{1}{4},\tfrac{1}{4},\tfrac{1}{4}\big)$, with $\Gamma(\frac14)^{-4}\simeq1/173$. We have checked~\eqref{eq:dirichlet} against an explicit enumeration of partitions. Since $R_{j1}^2=1/4$ for every $j$, including $j=1$, the distribution is symmetric among the four outgoing sectors and is the same for any incoming flavour, lepton or quark. This symmetry is special to $N_f=4$. For $N_f>4$, in the basis of the $SU(2)_M$ doublets, the sector of the incoming doublet has parameter $(1-\frac{2}{N_f})^2$ and each of the others $4/N_f^2$, and the dominant component of the outgoing state is then the $SU(2)_M$ partner of the incoming fermion and not its helicity-flipped state.

The parameters of the distribution are the mean fractions, and because they are smaller than one the density~\eqref{eq:dirichlet} diverges whenever any $x_j\to0$. The weight therefore sits at the corners of the simplex, where one share is close to one and the others close to zero. This, and not merely the size of the fluctuations, is why one sector typically carries most of the energy, as the upper block of Table~\ref{Tab:shares} shows. The mechanism is the one just described: the sector that receives the hardest boson quantum has most of the energy, and the second quantum lands in a different sector three times out of four. Which sector dominates is random, each of the four with probability $1/4$.

\begin{table}[ht]
\begin{equation} \nonumber
\begin{array}{l|cccc}
\text{share of }E &\ \text{mean}\ &\ \text{median}\ &\ P(>\tfrac12)\ &\ P(>0.9)\
\\\hline
\text{\it The four shares in order of size} & & & &
\\
\quad\text{the largest} & 0.72 & 0.73 & 0.88 & 0.22
\\
\quad\text{the second largest} & 0.21 & 0.20 & 0 & 0
\\
\quad\text{the two largest together} & 0.93 & 0.97 & 1 & 0.74
\\\hline
\text{\it Lepton and hadron shares} & & & &
\\
\quad\text{the lepton sector, }x & 0.25 & 0.09 & 0.22 & 0.06
\\
\quad\text{the three quark sectors together, }y=1-x & 0.75 & 0.91 & 0.78 & 0.51
\end{array}
\end{equation}
\caption{The partition of the energy $E$ of the outgoing state among the four flavour sectors, from the distribution~\eqref{eq:dirichlet}. In the upper block the four shares are ranked by size in each event, whichever sectors they belong to. The lower block gives the shares of the lepton sector and of the three quark sectors together. The table holds for any incoming flavour, with $E$ the energy of the fermion that reaches the core. For an incoming electron the lepton sector is that of the outgoing electron, and the quark sectors are the three anti-quark sectors. For an incoming quark the lepton sector is that of the positron, and the quark sectors are the quark's own sector and two anti-quark sectors. The upper block is from $8\times10^6$ sampled events, except for the probabilities of the largest share, which are four times those of a given sector. The lower block follows from the Beta distributions~\eqref{eq:beta}.}
\label{Tab:shares}
\end{table}

At $r\gtrsim1/\Lqcd$ the three quark sectors are no longer separate, and what remains observable is the two-way split between the lepton and the hadrons. One of the four sectors is leptonic and three are quark sectors, whatever the incoming flavour. Since a sum of Dirichlet components is again Dirichlet-distributed, with the parameters added, the split follows exactly from~\eqref{eq:dirichlet}. The lepton fraction $x$ and the hadronic fraction $y=1-x$ have the probability densities
\begin{equation}
f(x)\,=\,\frac{x^{-3/4}\,(1-x)^{-1/4}}{\pi\sqrt2}\,,\qquad
g(y)\,=\,\frac{y^{-1/4}\,(1-y)^{-3/4}}{\pi\sqrt2}\,,
\label{eq:beta}
\end{equation}
on the interval from $0$ to $1$. These are the Beta distributions with parameters $(\frac{1}{4},\frac{3}{4})$ and $(\frac{3}{4},\frac{1}{4})$, and $g(y)=f(1-y)$. They give the lower block of Table~\ref{Tab:shares}: the lepton typically receives little of the energy and the hadrons most of it, and $P(y>1-\delta)\simeq0.90\,\delta^{1/4}$ for small $\delta$.

The distribution~\eqref{eq:dirichlet} is a property of the massless outgoing state. We assume that the massive dynamics at $r\gtrsim1/m$ then proceeds with the energies that the electron sector and the three quark sectors together have received. This treats these two energies as classical labels of decohered branches. 
The Hamiltonians $H_j$ of the massless sectors, defined in~\eqref{eq:Hj}, commute with each other and with the line, so the joint distribution is well defined.
That the massive dynamics does not recohere branches with different partitions is an assumption, motivated by the fact that the leptonic and the hadronic sectors interact only through the boundary, apart from electromagnetic effects and from weak interactions, which are negligible at distances $r\gg1/m_W$.
It is supported by the classical model of appendix~\ref{sec:classical}, where the sector energies change during the decay only by amounts of order the sector gaps.

Three consequences follow. First, the energies received by the electron sector and by the hadronic sectors together are, up to shifts of $O(\Lqcd)$, the budgets available to their decays. Second, thresholds are set by the whole of $E$, not by $E/4$: the $(B+L)$-violating channel can draw on the full energy when the fluctuation puts it in the hadronic sectors, and it does so with the steeply rising probability $P(y>m_p/E)$. Third, the two-body kinematics of~\eqref{eq:flipsig-e} is a fair description of the typical event, in the sense that the hadronic system carries essentially the full energy, while the outgoing electron typically carries a small share of the energy. What is new is the spread.

\section{Fermion masses and the decay of the twisted state}
\label{Sec:mass}

\subsection{The string and its decay}
\label{sec:tension}

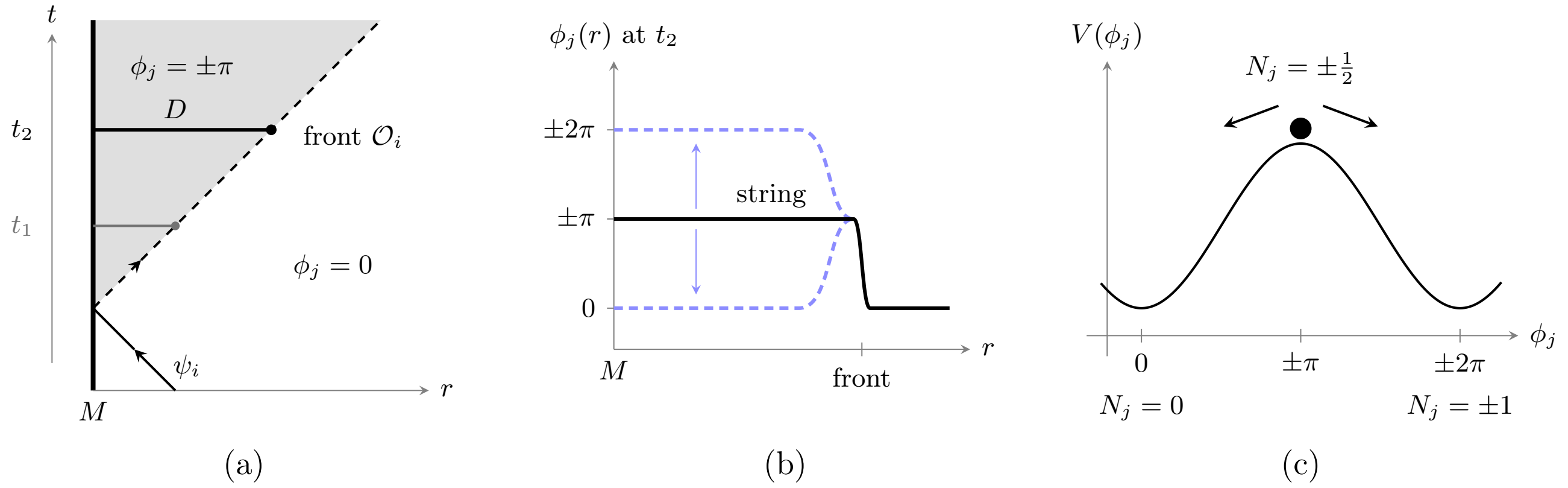
\begin{figure}[t]
\centering
\resizebox{\ifdim\width>\textwidth\textwidth\else\width\fi}{!}{%
\begin{tikzpicture}[>=stealth,scale=0.8,font=\footnotesize]
\begin{scope}[xshift=0cm]
\fill[gray!25] (0,1.0) -- (4.2,5.2) -- (0,5.2) -- cycle;
\draw[line width=1.6pt] (0,-0.2) -- (0,5.2);
\node[below] at (0,-0.2) {$M$};
\draw[->,gray] (0,-0.2) -- (4.9,-0.2) node[right,black] {$r$};
\draw[->,gray] (-0.6,0.2) -- (-0.6,5.0) node[above,black] {$t$};
\draw[->,thick] (1.2,-0.2) -- (0.6,0.4);
\draw[thick] (0.6,0.4) -- (0,1.0);
\node[right] at (1.0,0.15) {$\psi_i$};
\draw[->,thick,dashed] (0,1.0) -- (0.7,1.7);
\draw[thick,dashed] (0.7,1.7) -- (4.2,5.2);
\node[right] at (2.9,3.5) {front ${\cal O}_i$};
\draw[thick,black!55] (0,2.2) -- (1.2,2.2);
\fill[black!55] (1.2,2.2) circle (1.8pt);
\node[left,black!55] at (-0.7,2.2) {$t_1$};
\draw[very thick] (0,3.6) -- (2.6,3.6);
\fill (2.6,3.6) circle (2.2pt);
\node[left] at (-0.7,3.6) {$t_2$};
\node[above] at (1.2,3.6) {$D$};
\node at (1.3,4.5) {$\phi_j=\pm\pi$};
\node at (3.5,1.6) {$\phi_j=0$};
\node at (2.2,-1.3) {\normalsize (a)};
\end{scope}
\begin{scope}[xshift=7.6cm,yshift=0.6cm]
\draw[->,gray] (0,-0.2) -- (5.2,-0.2) node[right,black] {$r$};
\draw[gray] (-0.1,0.4) -- (0.1,0.4);
\draw[->,gray] (0,-0.2) -- (0,4.0) node[above,black] {$\phi_j(r)$ at $t_2$};
\draw[gray] (-0.1,1.7) -- (0.1,1.7);
\draw[gray] (-0.1,3.0) -- (0.1,3.0);
\node[left] at (-0.1,0.4) {$0$};
\node[left] at (-0.1,1.7) {$\pm\pi$};
\node[left] at (-0.1,3.0) {$\pm2\pi$};
\draw[very thick,blue!45,densely dashed] (0,3.0) -- (2.7,3.0) .. controls (3.2,3.0) and (3.1,1.7) .. (3.5,1.7);
\draw[very thick,blue!45,densely dashed] (0,0.4) -- (2.7,0.4) .. controls (3.2,0.4) and (3.1,1.7) .. (3.5,1.7);
\draw[->,blue!45] (1.2,1.85) -- (1.2,2.8);
\draw[->,blue!45] (1.2,1.55) -- (1.2,0.6);
\draw[very thick] (0,1.7) -- (3.5,1.7) .. controls (3.62,1.7) and (3.62,0.4) .. (3.74,0.4) -- (4.9,0.4);
\node[above] at (2.3,1.7) {string};
\draw[gray] (3.62,-0.3) -- (3.62,-0.1);
\node[below] at (3.62,-0.3) {front};
\node[below] at (0,-0.2) {$M$};
\node at (2.5,-1.9) {\normalsize (b)};
\end{scope}
\begin{scope}[xshift=15.3cm,yshift=2.2cm]
\draw[->,gray] (-0.8,-1.6) -- (5.5,-1.6) node[right,black] {$\phi_j$};
\draw[->,gray] (-0.5,-1.9) -- (-0.5,2.4) node[above,black] {$V(\phi_j)$};
\draw[thick,domain=-0.8:7.1,samples=120,smooth] plot ({0.74*\x},{-1.2*cos(\x r)});
\draw[gray] (0,-1.5) -- (0,-1.7);
\draw[gray] (2.325,-1.5) -- (2.325,-1.7);
\draw[gray] (4.65,-1.5) -- (4.65,-1.7);
\node[below] at (0,-1.7) {$0$};
\node[below] at (2.325,-1.7) {$\pm\pi$};
\node[below] at (4.65,-1.7) {$\pm2\pi$};
\fill (2.325,1.42) circle (4.5pt);
\draw[->,thick] (2.0,1.75) -- (1.2,1.45);
\draw[->,thick] (2.65,1.75) -- (3.45,1.45);
\node[below] at (0,-2.3) {$N_j=0$};
\node[below] at (4.65,-2.3) {$N_j=\pm1$};
\node[above] at (2.325,1.9) {$N_j=\pm\frac{1}{2}$};
\node at (2.325,-3.5) {\normalsize (c)};
\end{scope}
\end{tikzpicture}}
\caption{The twisted state in the massive theory. (a) The $(r,t)$ half-plane. The line $D$ runs from the monopole to the outgoing front, shown here on the equal-time slices $t_1$ and $t_2$, and sweeps the shaded region as the front moves out. In this region the mass term of every flavour has the opposite sign. (b) The profile of the bosonised field $\phi_j$ on the slice $t_2$. The front is a narrow step of height $\pm\pi$, and the string behind it is a plateau at $\phi_j=\pm\pi$. The dashed blue curves show the plateau after it has rolled to one of the two adjacent vacua, which is the breaking of the string. (c) The potential $V(\phi_j)=-\frac{m_j\Lambda}{\pi}\cos\phi_j$ of~\eqref{eq:mass}. The plateau of panel (b) sits at the maximum of the cosine potential, midway between two adjacent vacua, which correspond to the fermion numbers $N_j=0$ and $N_j=\pm1$ of the flavour $j$.}
\label{fig:line}
\end{figure}

The line $D=\exp(i\pi\sum_jN^{\rm out}_j)$ of section~\ref{sec:twisted} multiplies the outgoing fermion $\tilde\psi_j$ of every flavour by $-1$ and acts trivially on the incoming ones. A Dirac mass term $m_j\psi_j\tilde\psi_j^\dagger+{\rm h.c.}$ is therefore odd under $D$. The line is a symmetry line of the massless theory only, and once the masses are switched on it becomes a non-topological domain wall. The region swept by the line, between the monopole and the outgoing front, has the mass term of every flavour with flipped sign, as shown in Figure~\ref{fig:line}(a). On an equal-time slice the line is a string stretching radially from the core to the front. Its world-sheet is the shaded wedge of the Figure, and the edges of the world-sheet are the monopole world-line and the light-like trajectory of the front.

In bosonised form the semiton is the front, a narrow step in which $\phi_j$ changes by $\pm\pi$, as shown in Figure~\ref{fig:line}(b). The string is what the front leaves behind. All along it the field sits at $\phi_j=\pm\pi$, the {\it maximum} of the cosine potential $-\frac{m_j\Lambda}{\pi}\cos\phi_j$ of~\eqref{eq:mass}, in every flavour. This is Callan's observation~\cite{Callan:1983tm} that the half-solitons leave the field midway between two adjacent vacua of the cosine potential, which correspond to fermion numbers $0$ and $\pm1$ in that flavour, as in Figure~\ref{fig:line}(c).

Once the line is non-topological the string carries an energy per unit length, a tension. The flip of the sign of the mass term of flavour $j$ changes the Hamiltonian density by $2m_j\bar\psi_j\psi_j$, and the energy density of the string relative to the vacuum is the expectation value of this operator,
\begin{equation}
\sigma_j\,=\,2m_j\big|\langle\bar\psi\psi\rangle_j\big|\,=\,\frac{2m_j^2}{\pi}\ln\frac{\Lambda}{m_j}\,,
\label{eq:tension}
\end{equation}
where we used the condensate of a free massive fermion in two dimensions, $|\langle\bar\psi\psi\rangle_j|=\frac{m_j}{\pi}\ln\frac{\Lambda}{m_j}$, and $\Lambda$ is the ultraviolet cut-off on the fermion modes, the same scale as in~\eqref{eq:mass}. Its value in~\eqref{eq:tension} is set by the front and not by the monopole core. Modes with momenta well above the inverse width of the front, which is of order $E_j$, adjust adiabatically as the front passes and do not contribute to the energy of the string. To leading logarithmic accuracy the logarithm is therefore $\ln(E_j/m_j)$. The tension of the string is the sum $\sum_j\sigma_j$ over the flavours. In the bosonised picture the flip changes the potential by $\frac{2m_j\Lambda}{\pi}\cos\phi_j$. Classically this is the height $2m_j\Lambda/\pi$ of the maximum above the vacuum in Figure~\ref{fig:line}(c). In the quantum theory the fluctuations of $\phi_j$ reduce it by the factor $\langle\cos\phi_j\rangle=\frac{m_j}{\Lambda}\ln\frac{\Lambda}{m_j}$, which reproduces~\eqref{eq:tension}.

The string is stretched as the front moves out. The front in flavour $j$ carries the energy $E_j$, on average $E/4$ by~\eqref{eq:E4}, so the string could reach a length $E_j/\sigma_j\gg1/m_j$ before running out of energy. It does not get there, because the configuration $\phi_j=\pm\pi$ is unstable. The field along the string rolls off the maximum to one of the two adjacent vacua on the time scale $1/m_j$, as indicated in Figure~\ref{fig:line}(b), and the string breaks at a length of order $1/m_j$. Unlike the breaking of a confining string, this involves no tunnelling and no Schwinger-type exponential suppression.

\begin{figure}[t!]
\centering
\resizebox{\ifdim\width>\textwidth\textwidth\else\width\fi}{!}{%
\begin{tikzpicture}[>=stealth,scale=0.8,font=\footnotesize]
\newcommand{\breakaxes}{%
\draw[->,gray] (0,-0.5) -- (4.9,-0.5) node[right,black] {$r$};
\draw[->,gray] (0,-0.5) -- (0,3.9);
\foreach \y in {0,0.8,2.4,3.2} {\draw[gray,densely dotted] (0,\y) -- (4.5,\y);}
\draw[gray] (0,1.6) -- (4.5,1.6);
\foreach \y/\l in {0/{-2\pi},0.8/{-\pi},1.6/{0},2.4/{\pi},3.2/{2\pi}} {\draw[gray] (-0.1,\y) -- (0.1,\y); \node[left] at (-0.1,\y) {$\l$};}
\draw[gray] (3.3,-0.6) -- (3.3,-0.4);}
\begin{scope}[xshift=0cm,yshift=5.9cm]
\breakaxes
\node[above] at (0,3.9) {$\phi_e(r)$};
\draw[very thick] (0,0.8) -- (3.2,0.8) .. controls (3.3,0.8) and (3.3,1.6) .. (3.4,1.6) -- (4.5,1.6);
\fill (0,0.8) circle (2pt);
\node[below] at (1.6,0.8) {string};
\node[right] at (3.35,1.15) {$\frac{1}{2}\,e_L$};
\node at (2.3,4.8) {\normalsize (a) the string};
\end{scope}
\begin{scope}[xshift=7.6cm,yshift=5.9cm]
\breakaxes
\draw[very thick] (0,0) -- (2.3,0);
\draw[very thick,blue!70] (2.3,0) .. controls (2.7,0) and (2.7,0.8) .. (3.1,0.8) -- (3.2,0.8);
\draw[very thick] (3.2,0.8) .. controls (3.3,0.8) and (3.3,1.6) .. (3.4,1.6) -- (4.5,1.6);
\fill (0,0) circle (2pt);
\draw[->,black!65] (-1.25,0.72) -- (-1.25,0.12);
\node[left,black!65] at (-1.25,0.42) {$-\pi$};
\node[anchor=west,fill=white,inner sep=1.5pt] at (3.5,1.2) {$e_L$};
\node at (2.3,4.8) {\normalsize (b) helicity flip};
\end{scope}
\begin{scope}[xshift=15.2cm,yshift=5.9cm]
\breakaxes
\draw[very thick] (0,1.6) -- (2.3,1.6);
\draw[very thick,blue!70] (2.3,1.6) .. controls (2.7,1.6) and (2.7,0.8) .. (3.1,0.8) -- (3.2,0.8);
\draw[very thick] (3.2,0.8) .. controls (3.3,0.8) and (3.3,1.6) .. (3.4,1.6) -- (4.5,1.6);
\fill (0,1.6) circle (2pt);
\draw[->,black!65] (-1.25,0.88) -- (-1.25,1.48);
\node[left,black!65] at (-1.25,1.18) {$+\pi$};
\node[anchor=west,fill=white,inner sep=1.5pt] at (3.5,1.2) {$e_L$};
\node[anchor=east,blue!70,fill=white,inner sep=1.5pt] at (2.55,1.15) {soft $e^+$};
\node at (2.3,4.8) {\normalsize (c) $(B+L)$-violating};
\end{scope}
\begin{scope}[xshift=0cm,yshift=0cm]
\breakaxes
\node[above] at (0,3.9) {$\phi_q(r)$};
\draw[very thick] (0,2.4) -- (3.2,2.4) .. controls (3.3,2.4) and (3.3,1.6) .. (3.4,1.6) -- (4.5,1.6);
\fill (0,2.4) circle (2pt);
\node[above] at (1.6,2.4) {string};
\node[right] at (3.35,2.05) {$\frac{1}{2}\,\bar q$};
\node[below] at (0,-0.5) {$M$};
\node[below] at (3.3,-0.6) {front};
\end{scope}
\begin{scope}[xshift=7.6cm,yshift=0cm]
\breakaxes
\draw[very thick] (0,1.6) -- (2.3,1.6);
\draw[very thick,blue!70] (2.3,1.6) .. controls (2.7,1.6) and (2.7,2.4) .. (3.1,2.4) -- (3.2,2.4);
\draw[very thick] (3.2,2.4) .. controls (3.3,2.4) and (3.3,1.6) .. (3.4,1.6) -- (4.5,1.6);
\fill (0,1.6) circle (2pt);
\node[anchor=west,fill=white,inner sep=1.5pt] at (3.5,2.0) {$\bar q$};
\node[anchor=east,blue!70,fill=white,inner sep=1.5pt] at (2.55,2.05) {soft $q$};
\draw[->,black!65] (-1.25,2.32) -- (-1.25,1.72);
\node[left,black!65] at (-1.25,2.02) {$-\pi$};
\node[below] at (0,-0.5) {$M$};
\node[below] at (3.3,-0.6) {front};
\end{scope}
\begin{scope}[xshift=15.2cm,yshift=0cm]
\breakaxes
\draw[very thick] (0,3.2) -- (2.3,3.2);
\draw[very thick,blue!70] (2.3,3.2) .. controls (2.7,3.2) and (2.7,2.4) .. (3.1,2.4) -- (3.2,2.4);
\draw[very thick] (3.2,2.4) .. controls (3.3,2.4) and (3.3,1.6) .. (3.4,1.6) -- (4.5,1.6);
\fill (0,3.2) circle (2pt);
\draw[->,black!65] (-1.25,2.48) -- (-1.25,3.08);
\node[left,black!65] at (-1.25,2.78) {$+\pi$};
\node[anchor=west,fill=white,inner sep=1.5pt] at (3.5,2.0) {$\bar q$};
\node[below] at (0,-0.5) {$M$};
\node[below] at (3.3,-0.6) {front};
\end{scope}
\end{tikzpicture}}

\caption{The breaking of the string for an incoming $e_R$, in terms of the bosonised fields of Figure~\ref{fig:line}(b), with $\phi_j=0$ ahead of the front. The upper row is the electron sector of the outgoing state and the lower row is any one of its three quark sectors. (a) The twisted state~\eqref{eq:callan-e} before the string breaks. The fronts are half-steps of $+\pi$ and $-\pi$, and behind them the fields sit at the maxima of the cosine potential. (b) The helicity-flip channel, in which the boundary value at $r=0$ of every field shifts by $-\pi$. (c) The $(B+L)$-violating channel, in which it shifts by $+\pi$. The hard half-step of the front, in black, is the same in all panels. Behind it the soft field, in blue, makes a second half-step, which varies over the soft scale and brings the total charge of the sector to one unit or to zero. The curves show how the fermion number of the flavour is distributed in $r$, a step of $2\pi$ being one unit, and not which particles carry it. In a sector with a sizeable share of the energy the front contains one energetic fermion, $e_L$ or $\bar q$, in every panel. Where the total charge is zero this fermion is accompanied by its soft anti-particle (section~\ref{sec:hard}).}
\label{fig:breaking}
\end{figure}
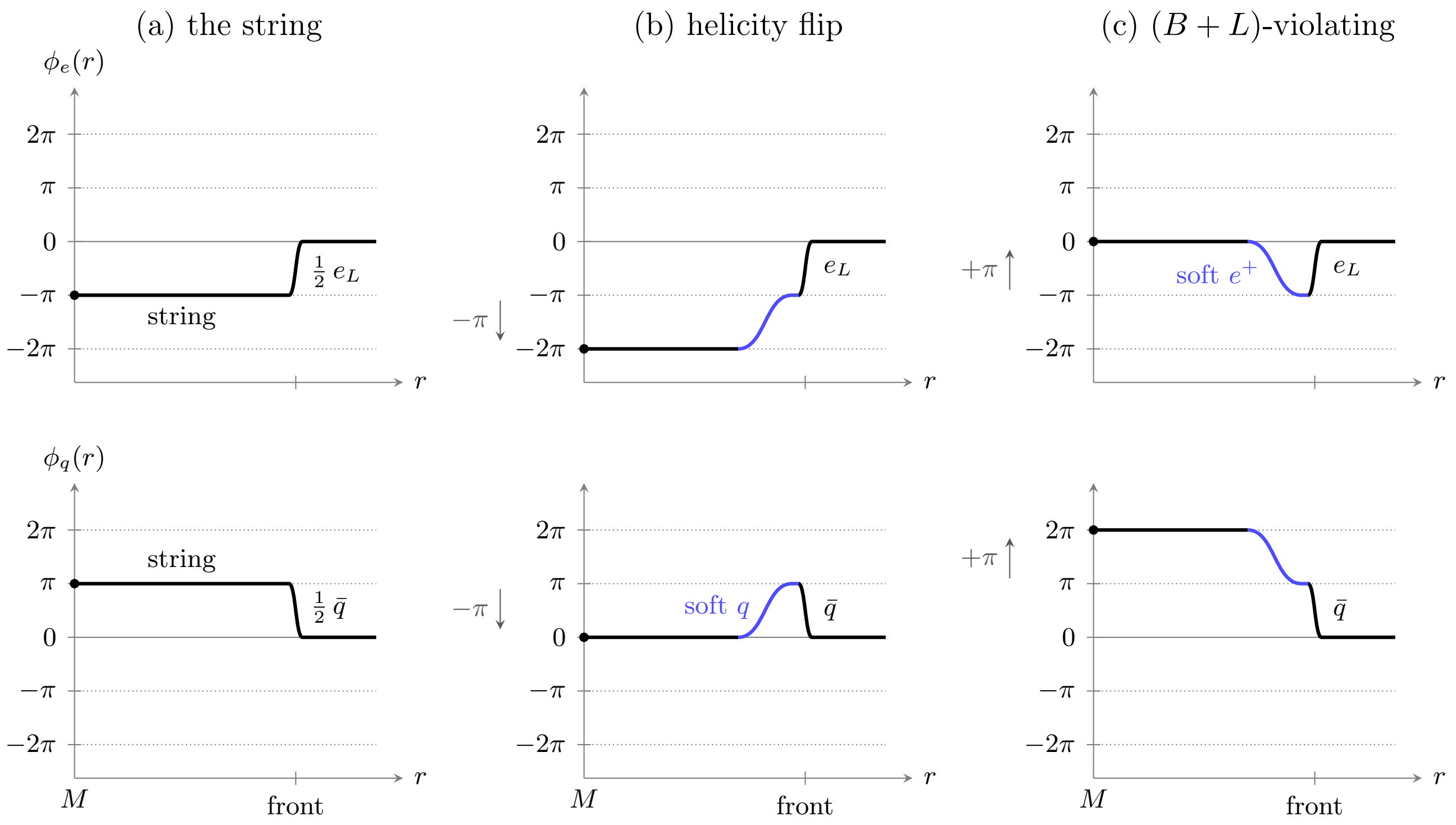

The breaking of the string is illustrated in Figure~\ref{fig:breaking} for an incoming $e_R$. Panel (a) shows the twisted state on the right-hand side of~\eqref{eq:callan-e}, with the string still intact. When the string breaks, the field along it moves by $\pm\pi$ to a neighbouring vacuum. A change of the field by $\pi$ is half a unit of charge, and it has to be accounted for at each end of the string. At the outer end the half-unit brings the total charge of the sector to an integer. It either adds to the half-unit of the front, which gives one unit, or cancels it. What this means for the particle content of the sector is discussed in section~\ref{sec:hard}. At the inner end the boundary value of the field shifts by $\pm\pi$, which is the monopole absorbing half a unit of charge in every flavour. Soft kinks and anti-kinks can be left in between. The energy released is $\sigma_jL$ with $L\sim1/m_j$, which is of order $m_j\ln(E_j/m_j)$ and small compared with $E_j$. The secondaries are therefore soft, with energies of order $m_j$, while the front keeps its $E_j$.

For the quark sectors the current quark mass is not the relevant scale. The condensate is that of QCD, the energy density of the flipped region is set by $m_q|\langle\bar qq\rangle|\sim m_\pi^2f_\pi^2$, and the sectors are gapped at the confinement scale. Their decay scale is $1/\Lqcd\sim1\,$fm, much shorter than the $1/m_e$ of the electron sector. As shown in section~\ref{sec:high}, the quark sectors therefore drive the decay, including the helicity flip $e_R\to e_L$.

In every flavour the field must thus roll to one of its two neighbouring vacua. By itself this would allow $2^4$ integer final states. Which of them can occur is fixed by the conservation laws, to which we now turn.

\subsection{The selection rule}
\label{sec:selrule}

Let $N_j$ be the fermion number of the Dirac flavour $j$, counting incoming and outgoing components alike (electron number, $u^1$-number, and so on). A Dirac mass term conserves every $N_j$. The operator $m_j\psi_j\tilde\psi_j^\dagger+{\rm h.c.}$ converts an incoming fermion into an outgoing one of the same flavour, or creates a fermion--anti-fermion pair of that flavour with one member incoming and the other outgoing, and neither changes $N_j$. The boundary does not conserve them. If over the whole process it absorbs incoming charges $a\in\Z^4$, it emits outgoing charges $Ra$, so by~\eqref{eq:R} it changes the flavour numbers by
\begin{equation}
\Delta N\,=\,(R-\one)\,a\,=\,-\frac{1}{2}\Big(\sum_{k=1}^{4} a_k\Big)\,(1,1,1,1)\,,
\label{eq:selrule}
\end{equation}
the same amount in every flavour. Starting from an incoming $\psi_i$, the final flavour content of any process, however many mass insertions it involves, is therefore
\begin{equation}
F^{\rm out}\,=\,\hat e_i\,+\,n\,(1,1,1,1)\,,\qquad n\in\Z\,,
\label{eq:selrule2}
\end{equation}
where $n$ is the number of units of $(1,1,1,1)$ that the monopole core has added to the flavour numbers over the whole process. It is an integer, since the final states of the massive theory have integer fermion numbers.
One unit is the outgoing set $e_L\,u_R^{1}\,u_R^{2}\,d_L^{\,3}$, an electron together with the quarks of a proton, so $n$ counts the $(B+L)$ violation, $\Delta B=\Delta L=n$. It is fixed through~\eqref{eq:selrule} by the total incoming charge absorbed by the boundary, $n=-\frac{1}{2}\sum_ka_k$. A single massless passage has $\sum_ka_k=1$, formally $n=-\frac{1}{2}$, and gives the twisted state. An integer final state requires the mass terms to send charge back to the core, so that $\sum_ka_k$ is even.

The mass terms preserve each $U(1)_j$ and the boundary violates them only uniformly, so the differences $N_j-N_k$ are exactly conserved. This is the conservation-law constraint on Callan's final states emphasised in~\cite{Sen:1983yq,Sen:1984qe}. In particular $Q_{el}$ and $B-L$, which are combinations of these differences, are conserved.

Of the $2^4$ states of section~\ref{sec:tension} only two are compatible with~\eqref{eq:selrule2}. They are the twisted charges $R\hat e_i$ shifted by $\pm\frac{1}{2}(1,1,1,1)$, with all four fields rolling in the same direction, and they have $n=0$ and $n=-1$.\footnote{Other values of $n$ are not excluded by the conservation laws, but they are not reached by the decay of the string. They would require the fields to move past the neighbouring vacua and over a further maximum of the potential, for which a field that rolls off a maximum and radiates does not have the energy.}
 For $e_R$ on $M$ they are
\begin{eqnarray}
n=0\,: && \makebox[6.0cm][l]{$e_R+M\,\to\,M+e_L\,,$} \,\,\,\, \textstyle\sum_ka_k=0\,, \ \ \Delta B=\Delta L=0\,,
\label{eq:n0}\\
n=-1\,: && \makebox[6.0cm][l]{$e_R+M\,\to\,M+(\overline{u})_{L1}(\overline{u})_{L2}(\overline{d}\,)_{R3}\,,$} \quad
\textstyle\sum_ka_k=2\,, \ \ \Delta B=\Delta L=-1\,.
\label{eq:nm1}
\end{eqnarray}
The process~\eqref{eq:n0} is the pure helicity flip, and~\eqref{eq:nm1} is the $(B+L)$-violating process. In both, all the final-state fermions are outgoing $s$-wave modes.\footnote{A three-fermion final state with the flavour numbers of~\eqref{eq:nm1} was proposed in~\cite{Csaki:2021ozp} as an alternative to the semiton state of the massless theory. It is a different state, $(\overline{u})_{L1}(\overline{u})_{L2}(\overline{d}\,)_{L3}$ in our notation, with the opposite chirality of the down anti-quark. A fermion of this chirality is an incoming mode in the $j_0$-wave. As an outgoing particle it must be in a higher partial wave, and it cannot emerge from the monopole core~\cite{Khoze:2023kiu}. In~\eqref{eq:nm1} all three anti-fermions are outgoing $s$-wave modes, which is possible only because the mass terms violate the chirality selection rule of the massless theory~\cite{Rubakov:1988aq}.}

\begin{figure}[t]
\centering
\resizebox{\ifdim\width>\textwidth\textwidth\else\width\fi}{!}{%
\begin{tikzpicture}[>=stealth,scale=0.8,font=\footnotesize]
\begin{scope}[xshift=0cm]
\node at (3.4,-1.2) {\normalsize (a) helicity flip};
\draw[line width=1.6pt] (0,-0.2) -- (0,5.2);
\node[below] at (0,-0.2) {$M$};
\draw[->,gray] (0,-0.2) -- (4.4,-0.2) node[right,black] {$r$};
\draw[->,thick] (1.4,-0.2) -- (0.7,0.5);
\draw[thick] (0.7,0.5) -- (0,1.2);
\node[right] at (1.2,0.15) {$e_R$};
\draw[->,thick,dashed] (0,1.2) -- (0.7,1.9);
\draw[thick,dashed] (0.7,1.9) -- (1.4,2.6);
\node[below right] at (0.65,1.85) {$\frac{1}{2}\big(e_L\,(\overline{u})_{L1}(\overline{u})_{L2}(\overline{d}\,)_{R3}\big)$};
\fill (1.4,2.6) circle (2.2pt);
\node[right] at (1.55,2.55) {$m_u$};
\draw[->,thick] (1.4,2.6) -- (0.7,3.3);
\draw[thick] (0.7,3.3) -- (0,4.0);
\node[above right,inner sep=1.5pt] at (0.7,3.3) {$(\overline{u})_{R1}$};
\draw[->,thick,dashed] (1.4,2.6) -- (2.0,3.2);
\draw[thick,dashed] (2.0,3.2) -- (2.5,3.7);
\node[right] at (2.35,3.15) {$\frac{1}{2}\big(e_L\,u^{1}_{R}\,(\overline{u})_{L2}(\overline{d}\,)_{R3}\big)$};
\draw[->,thick,dashed] (0,4.0) -- (0.5,4.5);
\draw[thick,dashed] (0.5,4.5) -- (0.9,4.9);
\node[right] at (0.85,4.45) {$\frac{1}{2}\big(e_L\,(\overline{u})_{L1}\,u^{2}_{R}\,d^{\,3}_{L}\big)$};
\draw[thick,decorate,decoration={brace,amplitude=5pt}] (6.4,4.75) -- (6.4,2.85);
\draw[->,thick] (6.75,3.85) -- (7.65,4.75);
\node[anchor=south,inner sep=1.5pt] at (7.7,4.85) {$e_L$};
\end{scope}
\begin{scope}[xshift=10.2cm]
\node at (4.0,-1.2) {\normalsize (b) $(B+L)$-violating};
\draw[line width=1.6pt] (0,-0.2) -- (0,5.2);
\node[below] at (0,-0.2) {$M$};
\draw[->,gray] (0,-0.2) -- (4.4,-0.2) node[right,black] {$r$};
\draw[->,thick] (1.4,-0.2) -- (0.7,0.5);
\draw[thick] (0.7,0.5) -- (0,1.2);
\node[right] at (1.2,0.15) {$e_R$};
\draw[->,thick,dashed] (0,1.2) -- (0.7,1.9);
\draw[thick,dashed] (0.7,1.9) -- (1.4,2.6);
\node[below right] at (0.65,1.85) {$\frac{1}{2}\big(e_L\,(\overline{u})_{L1}(\overline{u})_{L2}(\overline{d}\,)_{R3}\big)$};
\fill (1.4,2.6) circle (2.2pt);
\node[right] at (1.55,2.55) {$m_e$};
\draw[->,thick] (1.4,2.6) -- (0.7,3.3);
\draw[thick] (0.7,3.3) -- (0,4.0);
\node[above right,inner sep=1.5pt] at (0.7,3.3) {$e_R$};
\draw[->,thick,dashed] (1.4,2.6) -- (2.0,3.2);
\draw[thick,dashed] (2.0,3.2) -- (2.5,3.7);
\node[right] at (2.35,3.15) {$\frac{1}{2}\big((\overline{e})_{R}\,(\overline{u})_{L1}(\overline{u})_{L2}(\overline{d}\,)_{R3}\big)$};
\draw[->,thick,dashed] (0,4.0) -- (0.5,4.5);
\draw[thick,dashed] (0.5,4.5) -- (0.9,4.9);
\node[right] at (0.85,4.45) {$\frac{1}{2}\big(e_L\,(\overline{u})_{L1}(\overline{u})_{L2}(\overline{d}\,)_{R3}\big)$};
\draw[thick,decorate,decoration={brace,amplitude=5pt}] (7.35,4.75) -- (7.35,2.85);
\draw[->,thick] (7.7,3.85) -- (8.6,4.75);
\node[anchor=south east,inner sep=1.5pt] at (8.85,4.85) {$(\overline{u})_{L1}(\overline{u})_{L2}(\overline{d}\,)_{R3}$};
\end{scope}
\end{tikzpicture}}
\caption{Charge bookkeeping in the $(r,t)$ half-plane for the two processes~\eqref{eq:n0} and~\eqref{eq:nm1} at lowest order in the masses. Solid lines carry integer charges and dashed lines the half-integer charges of the twisted sector. In both panels the core first emits the twisted state~\eqref{eq:callan-e}. (a) The helicity-flip process~\eqref{eq:n0}. A mass insertion turns the half-unit of $(\overline{u})_{L1}$ into an incoming anti-quark $(\overline{u})_{R1}$, leaving half a unit of $u^1_R$ behind, and the core re-emits the anti-quark as a second twisted state. In the two outgoing states the half-units of quarks and anti-quarks of each flavour annihilate into radiation, and the two half-units of $e_L$ combine into the final $e_L$. (b) The anti-proton synthesis~\eqref{eq:nm1}. A mass insertion turns the half-unit of $e_L$ into an incoming $e_R$, leaving half a unit of $(\overline{e})_R$ behind, and the core re-emits the electron as a second twisted state. The half-units of $e_L$ and $(\overline{e})_R$ annihilate into radiation, and the half-units of the anti-quarks combine into the final $(\overline{u})_{L1}(\overline{u})_{L2}(\overline{d}\,)_{R3}$.}
\label{fig:bookkeeping}
\end{figure}
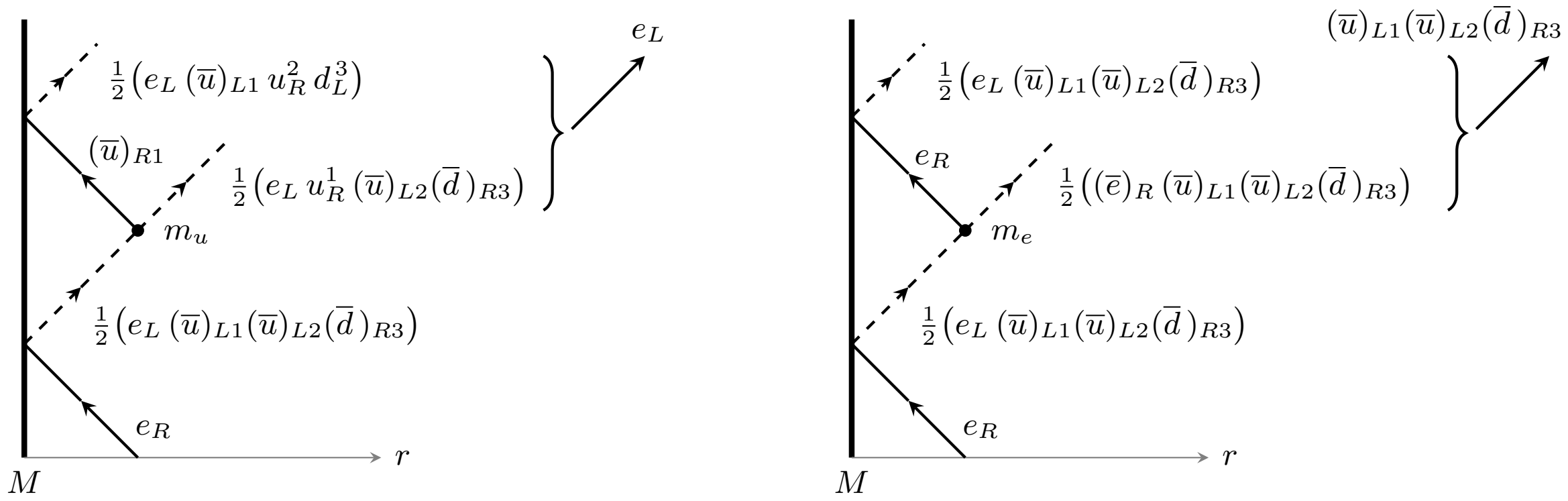

The bookkeeping for the two processes is illustrated in Figure~\ref{fig:bookkeeping}. In general, the core first emits the twisted state with charges $R\hat e_i$. At $r\sim1/m_j$ a mass insertion of flavour $j$ turns one unit of charge $q\,\hat e_j$, with $q=\pm1$, back towards the core, which re-emits it as $q\,R\hat e_j$. The total outgoing charge is
\begin{equation}
R\hat e_i\,-\,q\,\hat e_j\,+\,q\,R\hat e_j\,=\,\hat e_i\,-\,\frac{1}{2}(1+q)\,(1,1,1,1)\,,
\label{eq:bookkeeping}
\end{equation}
for any flavour $j$. A returned anti-fermion, $q=-1$, gives the helicity flip~\eqref{eq:n0}, as in panel (a) with $j=u^1$. A returned fermion, $q=+1$, gives the $(B+L)$-violating process~\eqref{eq:nm1}, as in panel (b) with $j=e$. Figure~\ref{fig:bookkeeping} shows the choices of $j$ for which all intermediate charges are $\pm\frac{1}{2}$.

Figure~\ref{fig:bookkeeping} shows only the lowest order in the masses. The processes~\eqref{eq:n0} and~\eqref{eq:nm1} do not rely on it. They follow from the conservation laws behind~\eqref{eq:selrule2}, which hold for any values of the masses and involve no expansion in $m_j$. Both final states exist only in the massive theory, and their hard production near the core is small. The amplitude with one mass insertion is of first order in $m_j$ and by dimensional analysis contributes a probability $\sim m_j^2/E^2$. The $O(1)$ conversion of the twisted state into integer fermions is the breaking of the string at $r\sim1/m_j$ described in section~\ref{sec:tension}, which is not captured at any finite order in the masses.

\subsection{Physical channels}
\label{sec:channels}

The results of this section combine into a simple picture. Each $U(1)_j$ is conserved in the bulk, so the half-units of charge of the twisted state can only be disposed of through the boundary. When the string breaks, the monopole takes up half a unit of charge in every flavour, and by the selection rule the sign is the same for all flavours. The boundary thus absorbs $\pm\frac{1}{2}(1,1,1,1)$, the charge of the twist operator $T$ in~\eqref{eq:Tproc1}. The decay of the twisted state is the line retracting into the monopole, which takes up its half-unit of charge with either sign. The two signs are indistinguishable in the massless theory, where $D$ is a $\Z_2$ line. They differ once the charges must be integer, and they leave the bulk with the $n=0$ or the $n=-1$ member of~\eqref{eq:selrule2}. There are thus exactly two outcomes for every incoming fermion,
\begin{subequations}
\label{eq:channels}
\begin{align}
e_R+M\ &\to\ \left\{
\begin{array}{l}
M+e_L+X_0\\[2pt]
M+(\overline{u})_{L1}(\overline{u})_{L2}(\overline{d}\,)_{R3}+X_0
\end{array}\right.
\label{eq:channels-e}\\[6pt]
u^{\,1}_L+M\ &\to\ \left\{
\begin{array}{l}
M+u^{\,1}_R+X_0\\[2pt]
M+(\overline{u})_{L2}(\overline{d}\,)_{R3}\,(\overline{e})_{R}+X_0
\end{array}\right.
\label{eq:channels-u}\\[6pt]
d^{\,3}_R+M\ &\to\ \left\{
\begin{array}{l}
M+d^{\,3}_L+X_0\\[2pt]
M+(\overline{u})_{L1}(\overline{u})_{L2}\,(\overline{e})_{R}+X_0
\end{array}\right.
\label{eq:channels-d}
\end{align}
\end{subequations}
and $u_L^{\,2}$ as $u_L^{\,1}$ with $1\leftrightarrow2$. Here $X_0$ is neutral in every conserved charge (lepton pairs, mesons, photons). In each case the upper line is the helicity flip, which conserves $B+L$, and the lower line is the $(B+L)$-violating channel, with $\Delta B=\Delta L=-1$. For $e_R$ the two lines of~\eqref{eq:channels-e} are the processes~\eqref{eq:n0} and~\eqref{eq:nm1} with the neutral radiation included, and the three anti-quarks have the quantum numbers of an anti-proton. After hadronisation they appear as an anti-proton, or as an anti-neutron accompanied by mesons of total electric charge $-1$, for example $\bar n\,+\,\pi^-$. In the second case the hadronisation has created a $d\bar d$ pair, which carries no conserved charge and belongs to $X_0$. The anti-neutron thus requires at least one additional pion and a hadronic energy above $m_n+m_\pi\simeq1.08\,\GeV$, so that close to the threshold the anti-nucleon is an anti-proton. Well above the threshold both occur, and their relative rate is a question of hadronisation that is not addressed here. This is the complete channel list, in agreement with what was anticipated in section~6 of~\cite{Khoze:2024hlb} on the basis of Callan's statistical rule. 
The selection rule of section~\ref{sec:selrule} is what restricts the statistical rule to these two channels.

In each of the processes~\eqref{eq:channels} the two channels differ only by the sign of the charge $\pm\frac{1}{2}(1,1,1,1)$ taken up by the monopole, which is common to all four flavours. The branching problem is therefore a single number per incoming fermion,
\begin{equation}
\br_{LR}(E)\,+\,\br_{B+L}(E)\,=\,1\,,
\end{equation}
where $\br_{LR}$ is the branching ratio of the helicity flip and $\br_{B+L}$ that of the $(B+L)$-violating channel.
The channel fixes the total charge of each sector. It does not fix how the charge and energy of a sector are divided between the hard front and the soft secondaries. The next section determines the branching ratios and then returns to this division.

\section{Branching ratios}
\label{Sec:br}

\subsection{High energy, $E\gg m_p$: a symmetry}
\label{sec:high}

Two questions arise when the string breaks (see section~\ref{sec:tension} and Figure~\ref{fig:breaking}). The first is decided at the inner end of the string, at $r=0$. The monopole absorbs the charge $\pm\frac{1}{2}(1,1,1,1)$, and the sign fixes the channel, the helicity flip of Figure~\ref{fig:breaking}(b) or the $(B+L)$-violating channel of Figure~\ref{fig:breaking}(c). The second question concerns the outer end of the string. The soft field behind the front fixes the total charge of each flavour sector, for example the electron number of the electron sector. This does not determine whether the final state contains an energetic particle of that flavour, such as a hard $e_L$, or only soft ones. It is shown in section~\ref{sec:hard} that this is already fixed by the outgoing state of the massless theory, before the string breaks, and in the same way in both channels. The rest of this subsection concerns the decision at the monopole.

The twisted state lies exactly midway between the two integer final states. Its charges are
\begin{equation}
R\hat e_i\,=\,\hat e_i\,-\,\tfrac{1}{2}(1,1,1,1)\,=\,\big[\hat e_i-(1,1,1,1)\big]\,+\,\tfrac{1}{2}(1,1,1,1)\,,
\label{eq:midway}
\end{equation}
so the same operator can be written in two ways. For an incoming electron,
\begin{equation}
{\cal O}_e\,=\,T\,e_L\,=\,T^\dagger\,(\overline{u})_{L1}(\overline{u})_{L2}(\overline{d}\,)_{R3}\,,
\label{eq:twoways}
\end{equation}
where $T^\dagger=\frac{1}{2}\big(u_R^{1}\,u_R^{2}\,d_L^{\,3}\,e_L\big)$ is the conjugate of the twist operator $T$ of~\eqref{eq:Tproc1} and carries the charge $+\frac{1}{2}(1,1,1,1)$. Each side is a single vertex operator, and the equality is that of the flavour vectors in~\eqref{eq:midway}. In the second form the half-unit of each quark in $T^\dagger$ cancels half a unit of the corresponding anti-quark, in the way a quark--anti-quark pair annihilates, and half a unit of $e_L$ is left. Both $T$ and $T^\dagger$ can sit at the end of the same line $D$, because $D$ is a $\Z_2$ line, $D=D^{-1}$, and has no orientation. When the string breaks, the monopole takes up either the charge of $T$, which leaves the bulk with the charges of $e_L$, the helicity flip, or the charge of $T^\dagger$, which leaves it with those of the three anti-quarks, the $(B+L)$-violating state. In both cases the hard front is the same operator ${\cal O}_e$, and the difference is carried by the soft fields behind it.

The operators $T$ and $T^\dagger$ are exchanged by the charge conjugation $C$ of the two-dimensional theory, which reverses all the flavour numbers and is the $CP$ transformation in four dimensions. It is a symmetry of the bulk theory with its mass terms and strong interactions, of the boundary condition~\eqref{eq:R}, which is linear in the fermion-number densities, and of the line $D$. The outgoing state itself is not invariant, since $C$ maps ${\cal O}_e$ into its conjugate, and the symmetry will be applied below to the string alone.

At high energy the hard fronts do not influence how the string breaks. The front of flavour $j$ carries the energy $E_j\gg m_j$ and has the width $1/E_j$, while the string varies and decays on the scale $1/m_j$. The front is insensitive to the mass term up to $O(m_j/E_j)$, and the soft modes see it only as the place where the line ends. The decay is then a problem for the string alone. The string is its own conjugate, and $C$ exchanges its two ways of breaking. They therefore have equal probability,
\begin{equation}
\br_{LR}\,=\,\br_{B+L}\,=\,\frac{1}{2}\,, \qquad{\rm for}\,\,E\gg m_p\,.
\label{eq:half}
\end{equation}
This is the equal weighting of Callan's statistical rule~\cite{Callan:1983tm,Rubakov:1988aq} and the high-energy limit of~\eqref{eq:flipsig-e}, now derived. Appendix~\ref{app:reflection} repeats the argument in bosonised form, where $C$ is a reflection of the soft fields about the maximum of the cosine potential.

The result is exact as a limit, not at finite $E$. The fronts carry the charges $R\hat e_i$ and are not invariant under $C$, and their $O(m_j/E_j)$ coupling to the soft modes, which was dropped, breaks the symmetry. In the classical version of the model this coupling decides the channel deterministically. Equation~\eqref{eq:half} therefore rests on the quantum fluctuations of the soft modes dominating this bias for $E\gg m_j$, which is estimated in section~\ref{sec:quantum} but is not a theorem.

The argument also requires the twist charge to be exactly $\frac{1}{2}$, as it is for $M$ with one generation of fermions, $N_f=4$. For a twist charge $2/N_f$ with $N_f>4$ the line has an orientation, the string is not its own conjugate, and the twisted charges are no longer midway between two integer states. The two channels are then not related by a symmetry, and their branching ratios are not fixed by the present argument. This is the case for the higher monopoles of~\cite{Khoze:2024hlb}.

The sectors need not decay at the same scale. For $M$ the quark sectors decay at $1\,$fm and the electron sector only at $1/m_e$. The string is broken by the quark sectors at $t\sim1\,$fm, when the electron sector is still massless. Since the monopole absorbs its half-unit of charge in all four flavours at once, it emits into the electron sector a second half-kink, of width $\sim1\,$fm and energy $O(\Lqcd)$, whose charge $\pm\frac{1}{2}$ makes the total charge of the electron sector integer. Charge conjugation is a symmetry for any values of the masses, so~\eqref{eq:half} stands, with corrections of order $\Lqcd/E_j$ from the quark fronts. 
Because the $E_j$ fluctuate according to~\eqref{eq:dirichlet}, the approach to~\eqref{eq:half} is governed in practice by the probability that the hadronic sectors receive enough energy (section~\ref{sec:threshold}).

\subsection{The energetic particles in the final state}
\label{sec:hard}

This subsection answers the second question of section~\ref{sec:high}: does the final state contain an energetic particle of a given flavour, for example a hard $e_L$? The choice between the two channels, the helicity flip and the $(B+L)$-violating one, is made at the monopole. In every flavour sector it shifts the field behind the front by $+\pi$ or by $-\pi$, which changes the total charge of the sector by $\pm\frac{1}{2}$. These are the blue half-steps in panels (b) and (c) of Figure~\ref{fig:breaking}. In the quark sectors the half-step is the field along the string rolling to the neighbouring vacuum. In the electron sector, which is still massless at $1\,$fm since $1/m_e\gg1\,$fm, it is the soft half-kink of section~\ref{sec:high}. In both cases the field varies over about $1\,$fm and has no components at momenta well above $\Lqcd$, so it cannot add or remove an energetic particle. The energetic particles of a sector, which we call its hard content, are therefore determined by the hard front alone, the black half-step of the figure. The front is the same in both channels, and so is the hard content, up to corrections of order $\Lqcd/E_j$.

The front itself does not consist of a definite number of fermions and anti-fermions, see e.g.~\cite{Tachikawa:2026cxd,Tong:2026apr}.
Its fermionic content can be computed exactly in the massless theory, in the box normalisation of section~\ref{sec:dirichlet}. The details are given in appendix~\ref{app:hard}. Consider a sector that carries half a unit of the fermion $f$. In~\eqref{eq:callan-e} $f$ is $e_L$ for the electron sector and an anti-quark for each of the quark sectors. The state of the sector is created by the vertex operator $V$ of charge $\frac{1}{2}$, and the occupation numbers of the fermion modes follow from the four-point function of $f^\dagger f$ with $V$ and its conjugate. In radial variables,
\begin{equation}
\big\langle V^\dagger(1)\,f^\dagger(z')\,f(z)\,V(\xi)\big\rangle\,=\,\frac{(1-\xi)^{-1/4}}{z'-z}\left[\frac{(1-z)\,(z'-\xi)}{(1-z')\,(z-\xi)}\right]^{1/2}\,.
\label{eq:fourpt}
\end{equation}
This is the standard correlator of four vertex operators of a free chiral boson, $\prod_{i<j}(z_i-z_j)^{q_iq_j}$ for charges $q_i$ at the points $z_i$~\cite{Ginsparg:1988ui,DiFrancesco:1997nk}, with $f^\dagger$, $f$ and $V$ of charges $+1$, $-1$ and $+\frac{1}{2}$. Here $V^\dagger$ and $V$ sit at the radii $1$ and $\xi<1$. The power of $\xi$ records the energy of the sector, so $\xi$ is the counting variable of section~\ref{sec:dirichlet}. The expansion in powers of $z$ and $z'$ gives the occupation numbers of the modes of $f$ and of its anti-particle, and the expansion in powers of $\xi$ resolves them according to that energy. The energies of the other three sectors are summed over with the weights of section~\ref{sec:dirichlet}, whose generating function is $(1-\xi)^{-3/4}$. This gives the mean occupation numbers exactly, for any total energy $E$ of the outgoing state. They are written out in~\eqref{eq:Npm} of appendix~\ref{app:hard}. In the continuum limit the sector contains on average $n_+(\omega|E)\,d\omega$ quanta of $f$ and $n_-(\omega|E)\,d\omega$ quanta of its anti-particle with energies between $\omega$ and $\omega+d\omega$, where
\begin{equation}
n_\pm(\omega|E)\,=\,\frac{1}{\pi^2E}\int_0^{E/\omega-1}dy\,\left[\frac{(1+y)\,(E/\omega-y)}{y\,(E/\omega-1-y)}\right]^{\pm1/2}\qquad(\omega<E)\,.
\label{eq:nspec}
\end{equation}
This formula was first obtained in~\cite{Tong:2026apr}, using a different method, from the fermion propagator in the outgoing state of a Lorentzian wavepacket, which is then resolved into components of fixed total energy. In our approach it arises as the continuum limit of the occupation numbers~\eqref{eq:Npm}. 
These, together with the four-point function~\eqref{eq:fourpt} and the expansions~\eqref{eq:AB} and~\eqref{eq:Cm} from which they are built, are derived here and hold exactly in the massless theory, for every total energy $E$.
 We have verified numerically, at $E=400$ in units of the level spacing, that~\eqref{eq:Npm} agrees with~\eqref{eq:nspec} (appendix~\ref{app:hard}). For $\omega\ll E$ both $n_+$ and $n_-$ behave as $1/(\pi^2\omega)$. This is a cloud of soft particle--anti-particle pairs, which in the massive theory considered here is cut off at $\omega\sim m_j$. The difference $n_+-n_-$ integrates to the charge of the sector, $\int_0^Ed\omega\,(n_+-n_-)=\frac{1}{2}$, and the energy is $\int_0^Ed\omega\,\omega\,(n_++n_-)=E/4$, in agreement with~\eqref{eq:E4}.

\begin{table}[ht]
\begin{equation} \nonumber
\begin{array}{l|ccccccc}
\text{cut }c &\ 0.05\ &\ 0.1\ &\ 0.2\ &\ 1/3\ &\ 0.5\ &\ 0.7\ &\ 0.9\
\\\hline
\text{quanta of }f\text{ above }cE & 0.57 & 0.47 & 0.37 & 0.28 & 0.19 & 0.10 & 0.03
\\
\text{anti-particles of }f\text{ above }cE & 0.12 & 0.07 & 0.03 & 0.01 & 0.003 & 0.0005 & 10^{-5}
\\
P(x>c) & 0.57 & 0.49 & 0.39 & 0.30 & 0.22 & 0.14 & 0.06
\end{array}
\end{equation}
\caption{The energetic particles in any one of the four flavour sectors of the outgoing state of total energy $E$. The sector carries half a unit of the fermion $f$, which is $e_L$ for the electron sector and an anti-quark for each quark sector of~\eqref{eq:callan-e}. The first two lines are the mean numbers of quanta of $f$ and of its anti-particle with energies above $cE$, given by the integrals of $n_+$ and $n_-$ of~\eqref{eq:nspec} from $cE$ to $E$. The last line is the probability that the sector holds more than the fraction $c$ of the energy, from~\eqref{eq:beta}. For $c\geq\frac{1}{2}$ at most one quantum can have an energy above $cE$, so the entries in the first two lines are then the probabilities that such a quantum is present.}
\label{Tab:hard}
\end{table}
Table~\ref{Tab:hard} compares the mean numbers of hard quanta in a sector, $\int_{cE}^{E}d\omega\,n_\pm(\omega|E)$, with the probability that the sector holds more than the fraction $c$ of the energy, $P(x>c)=\int_c^1dx\,\frac{x^{-3/4}(1-x)^{-1/4}}{\pi\sqrt2}$, which is the integral of the Beta density of~\eqref{eq:beta}. The number of hard fermions and this probability nearly coincide. A sector that has received a sizeable share of the energy thus contains, with a probability close to one, a single hard fermion that carries most of it, and a hard anti-fermion is rare. A sector with more than half of the energy contains a fermion above $E/2$ in $86\%$ of the events. In a sector with a sizeable share of the energy the hard half-step of Figure~\ref{fig:breaking} is therefore not half a particle, which the soft field would complete or remove. It is one energetic fermion together with a cloud of softer fermions and anti-fermions, and the cloud carries the net fermion number $-\frac{1}{2}$ that brings the total to $\frac{1}{2}$. The soft half-step of the figure adds $\pm\frac{1}{2}$ to the soft part of the sector and leaves the energetic fermion in place.

For the electron sector, with $E_j=xE$, the outcome is thus governed by the share $x$ of~\eqref{eq:beta}, and not by a second, independent choice. When $x$ is sizeable, a hard $e_L$ carrying most of $xE$ is present in either channel. In the helicity-flip channel it is the outgoing electron, and the soft part of the sector is neutral. In the $(B+L)$-violating channel the electron number of the sector is zero, and the hard $e_L$ is accompanied by a soft positron of energy $O(\Lqcd)$. When $x$ is small the whole sector is soft. A hard $e^+e^-$ pair that carries $xE$ without a hard net electron is rare. Since $P(x>0.1)=0.49$, a helicity-flipped electron with more than a tenth of the energy is present in about half of the events of each channel. The resulting final states are listed in~\eqref{eq:classes} of section~\ref{Sec:lepton}. For the quark sectors the hard quark content is not observable, and how often the anti-nucleon carries most of the hadronic energy is a question of hadronisation, which is not answered here.

\subsection{Low energy, $E\ll\Lqcd$: decoupling}
\label{sec:low}

The $(B+L)$-violating channel carries baryon number $-1$, and the lightest state with this baryon number is an anti-nucleon. For $E<m_p$ the channel is therefore closed by energy conservation and only the helicity flip remains. For $E\ll\Lqcd$ a stronger statement holds. The quark sectors are gapped at $1\,$fm by the chiral condensate, and a twisted state whose quark sectors carry at most $E\ll\Lqcd$ cannot propagate past $1\,$fm. The massless region is then not resolved, and the effective boundary condition at $r\sim1\,$fm involves the electron alone. This is the low-energy fixed point in the sense of~\cite{Polchinski:1984uw,Affleck:1993np,Maldacena:1995pq}. For a single Dirac fermion the only charge-conserving boundary condition is the vector-like one, $R=\one$ in the convention of~\eqref{eq:R}, which is the KYG helicity flip,
\begin{equation}
\br_{LR}\,=\,1\,,\qquad \sigma(e_R+M\to M+e_L)\,=\,\frac{\pi}{p^2}\qquad(E\ll\Lqcd)\,.
\label{eq:low}
\end{equation}
The quark sectors do not participate at all. The electron is neither partly reflected without flip nor partly absorbed into soft hadronic excitations, $X_0$ is empty, and the process is the two-body $e_R\to e_L$ with unit probability. It is made possible by the gap of the three quark sectors and not by $m_e$. In the non-relativistic limit $p=m_e\beta$, and~\eqref{eq:low} is the large cross-section $\pi/(m_e^2\beta^2)$ of~\cite{Khoze:2024hlb}.

\subsection{Threshold and interpolation}
\label{sec:threshold}

The $(B+L)$-violating channel needs an anti-baryon from the hadronic sectors, so the hadronic energy must satisfy $yE\ge m_p$. The threshold is $E=m_p$, as in~\eqref{eq:flipsig-e} and in~\cite{Candela:2025gwp}. It is reached when the fluctuation~\eqref{eq:dirichlet} puts the whole energy into the hadronic sectors, and not when the mean hadronic energy reaches $m_p$. 
The mean is $\frac{3}{4}E$, since three of the four sectors are quark sectors, and it equals $m_p$ at $E=\frac{4}{3}m_p$. This value of $E$ plays no special role, since the threshold is set by the events with the largest hadronic share, not by the average. By~\eqref{eq:beta} the hadronic sectors together receive at least $m_p$ with the probability $P(y>m_p/E)\simeq0.90\,(1-m_p/E)^{1/4}$ near threshold, which rises steeply and is already $0.5$ at $E=1.1\,m_p$. This is the probability that the $(B+L)$-violating channel is open. It is a property of the energy partition and not a branching ratio. Since the energy of a sector is fixed only up to a shift of order $\Lqcd$ (section~\ref{sec:dirichlet}), it is not meaningful for $E-m_p\lesssim\Lqcd$.

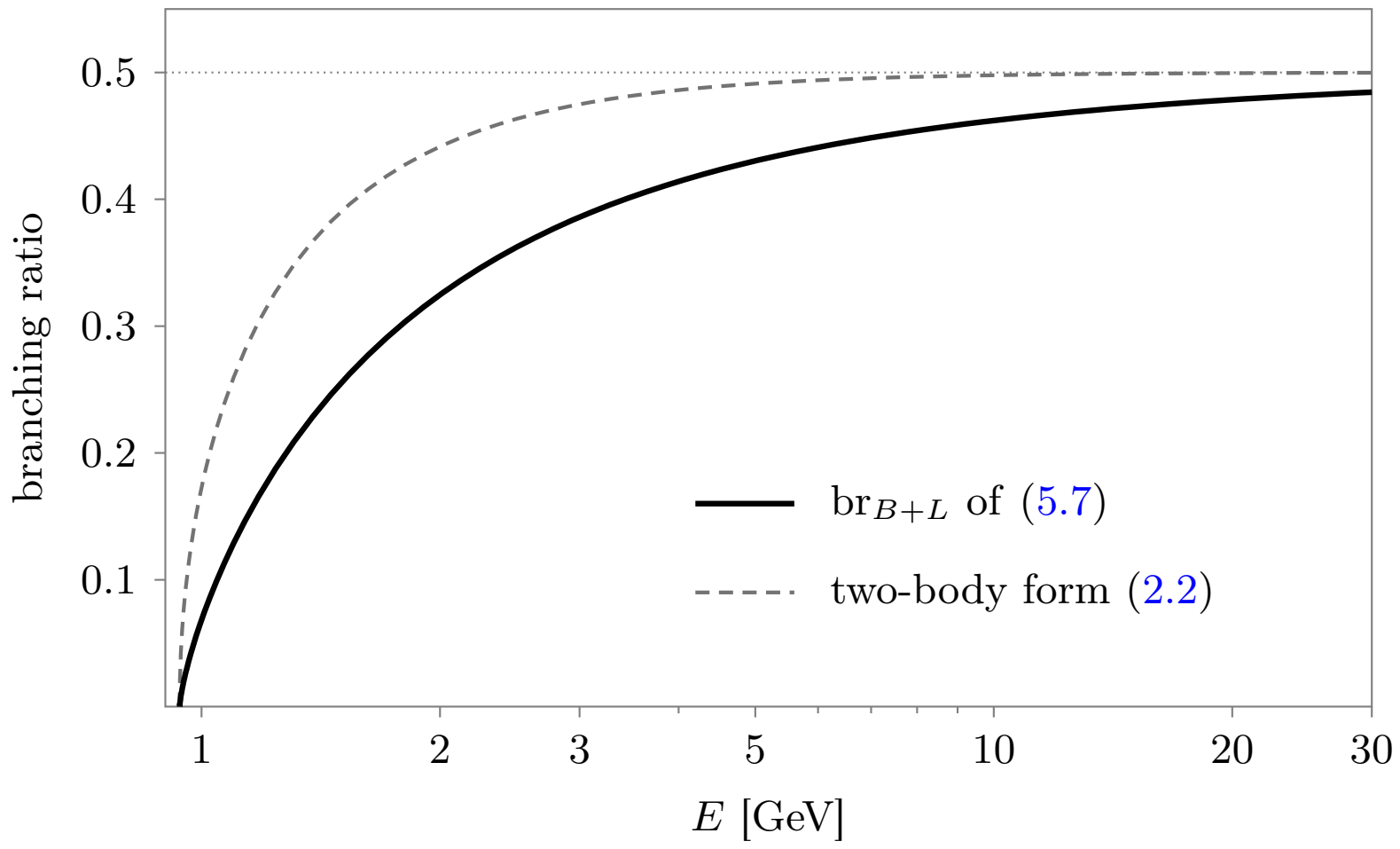
\begin{figure}[t]
\centering
\begin{tikzpicture}[>=stealth,font=\footnotesize]
\draw[gray] (0,0) rectangle (9.0,5.2);
\draw[gray] (0,0.945) -- (-0.08,0.945) node[left,black] {$0.1$};
\draw[gray] (0,1.891) -- (-0.08,1.891) node[left,black] {$0.2$};
\draw[gray] (0,2.836) -- (-0.08,2.836) node[left,black] {$0.3$};
\draw[gray] (0,3.782) -- (-0.08,3.782) node[left,black] {$0.4$};
\draw[gray] (0,4.727) -- (-0.08,4.727) node[left,black] {$0.5$};
\draw[gray] (0.270,0) -- (0.270,-0.08) node[below,black] {$1$};
\draw[gray] (2.049,0) -- (2.049,-0.08) node[below,black] {$2$};
\draw[gray] (3.090,0) -- (3.090,-0.08) node[below,black] {$3$};
\draw[gray] (4.401,0) -- (4.401,-0.08) node[below,black] {$5$};
\draw[gray] (6.180,0) -- (6.180,-0.08) node[below,black] {$10$};
\draw[gray] (7.959,0) -- (7.959,-0.08) node[below,black] {$20$};
\draw[gray] (9.000,0) -- (9.000,-0.08) node[below,black] {$30$};
\draw[gray] (3.829,0) -- (3.829,-0.05);
\draw[gray] (4.869,0) -- (4.869,-0.05);
\draw[gray] (5.265,0) -- (5.265,-0.05);
\draw[gray] (5.608,0) -- (5.608,-0.05);
\draw[gray] (5.910,0) -- (5.910,-0.05);
\draw[gray,densely dotted] (0,4.727) -- (9.0,4.727);
\draw[thick,densely dashed,black!55] (0.107,0) -- (0.107,0.067) -- (0.107,0.074) -- (0.107,0.082) -- (0.107,0.091) -- (0.107,0.101) -- (0.108,0.112) -- (0.108,0.124) -- (0.108,0.138) -- (0.108,0.153) -- (0.109,0.169) -- (0.109,0.188) -- (0.109,0.208) -- (0.110,0.231) -- (0.111,0.256) -- (0.112,0.284) -- (0.113,0.314) -- (0.114,0.348) -- (0.115,0.386) -- (0.117,0.428) -- (0.120,0.474) -- (0.123,0.525) -- (0.126,0.582) -- (0.131,0.645) -- (0.136,0.714) -- (0.143,0.790) -- (0.151,0.874) -- (0.162,0.966) -- (0.174,1.067) -- (0.189,1.178) -- (0.208,1.299) -- (0.230,1.431) -- (0.258,1.574) -- (0.291,1.729) -- (0.332,1.895) -- (0.381,2.072) -- (0.440,2.259) -- (0.510,2.455) -- (0.594,2.658) -- (0.694,2.865) -- (0.812,3.074) -- (0.819,3.086) -- (0.958,3.292) -- (1.097,3.466) -- (1.235,3.616) -- (1.374,3.745) -- (1.513,3.857) -- (1.651,3.955) -- (1.790,4.041) -- (1.929,4.116) -- (2.067,4.183) -- (2.206,4.242) -- (2.345,4.294) -- (2.483,4.340) -- (2.622,4.382) -- (2.761,4.418) -- (2.899,4.451) -- (3.038,4.480) -- (3.177,4.506) -- (3.315,4.529) -- (3.454,4.550) -- (3.593,4.568) -- (3.731,4.585) -- (3.870,4.600) -- (4.008,4.613) -- (4.147,4.625) -- (4.286,4.635) -- (4.424,4.645) -- (4.563,4.653) -- (4.702,4.661) -- (4.840,4.668) -- (4.979,4.674) -- (5.118,4.679) -- (5.256,4.684) -- (5.395,4.689) -- (5.534,4.693) -- (5.672,4.696) -- (5.811,4.699) -- (5.950,4.702) -- (6.088,4.705) -- (6.227,4.707) -- (6.366,4.709) -- (6.504,4.711) -- (6.643,4.713) -- (6.782,4.714) -- (6.920,4.716) -- (7.059,4.717) -- (7.198,4.718) -- (7.336,4.719) -- (7.475,4.720) -- (7.613,4.720) -- (7.752,4.721) -- (7.891,4.722) -- (8.029,4.722) -- (8.168,4.723) -- (8.307,4.723) -- (8.445,4.724) -- (8.584,4.724) -- (8.723,4.724) -- (8.861,4.725) -- (9.000,4.725);
\draw[very thick] (0.107,0) -- (0.107,0.005) -- (0.107,0.006) -- (0.107,0.007) -- (0.107,0.008) -- (0.107,0.010) -- (0.108,0.011) -- (0.108,0.013) -- (0.108,0.016) -- (0.108,0.018) -- (0.109,0.021) -- (0.109,0.025) -- (0.109,0.029) -- (0.110,0.034) -- (0.111,0.039) -- (0.112,0.046) -- (0.113,0.054) -- (0.114,0.063) -- (0.115,0.073) -- (0.117,0.085) -- (0.120,0.100) -- (0.123,0.116) -- (0.126,0.135) -- (0.131,0.158) -- (0.136,0.184) -- (0.143,0.214) -- (0.151,0.250) -- (0.162,0.290) -- (0.174,0.338) -- (0.189,0.392) -- (0.208,0.455) -- (0.230,0.528) -- (0.258,0.611) -- (0.291,0.705) -- (0.332,0.812) -- (0.381,0.933) -- (0.440,1.069) -- (0.510,1.219) -- (0.594,1.384) -- (0.694,1.564) -- (0.812,1.757) -- (0.819,1.769) -- (0.958,1.973) -- (1.097,2.158) -- (1.235,2.326) -- (1.374,2.480) -- (1.513,2.621) -- (1.651,2.750) -- (1.790,2.870) -- (1.929,2.981) -- (2.067,3.083) -- (2.206,3.178) -- (2.345,3.267) -- (2.483,3.349) -- (2.622,3.426) -- (2.761,3.497) -- (2.899,3.564) -- (3.038,3.627) -- (3.177,3.685) -- (3.315,3.740) -- (3.454,3.791) -- (3.593,3.839) -- (3.731,3.884) -- (3.870,3.927) -- (4.008,3.967) -- (4.147,4.005) -- (4.286,4.040) -- (4.424,4.074) -- (4.563,4.105) -- (4.702,4.135) -- (4.840,4.163) -- (4.979,4.190) -- (5.118,4.215) -- (5.256,4.239) -- (5.395,4.261) -- (5.534,4.283) -- (5.672,4.303) -- (5.811,4.322) -- (5.950,4.341) -- (6.088,4.358) -- (6.227,4.374) -- (6.366,4.390) -- (6.504,4.405) -- (6.643,4.419) -- (6.782,4.433) -- (6.920,4.445) -- (7.059,4.458) -- (7.198,4.469) -- (7.336,4.480) -- (7.475,4.491) -- (7.613,4.501) -- (7.752,4.511) -- (7.891,4.520) -- (8.029,4.529) -- (8.168,4.537) -- (8.307,4.545) -- (8.445,4.553) -- (8.584,4.560) -- (8.723,4.567) -- (8.861,4.574) -- (9.000,4.580);
\draw[very thick] (3.954,1.513) -- (4.692,1.513);
\node[anchor=west] at (4.826,1.513) {$\br_{B+L}$ of~\eqref{eq:interp}};
\draw[thick,densely dashed,black!55] (3.954,0.851) -- (4.692,0.851);
\node[anchor=west] at (4.826,0.851) {two-body form~\eqref{eq:flipsig-e}};
\node[below] at (4.5,-0.55) {$E$ [GeV]};
\node[rotate=90] at (-1.0,2.6) {branching ratio};
\end{tikzpicture}
\caption{The branching ratio of the $(B+L)$-violating channel for an incoming electron of energy $E$ in the rest frame of the monopole. The solid curve is the model~\eqref{eq:interp} and the dashed curve the two-body form~\eqref{eq:flipsig-e}. Both start at the threshold $E=m_p$ and approach $\frac{1}{2}$, shown by the dotted line.}
\label{fig:br}
\end{figure}

The analysis so far does not fix the shape of $\br_{B+L}(E)$ between~\eqref{eq:low} and~\eqref{eq:half}. In this range the hard-front approximation fails in the quark sectors and the boundary problem is strongly coupled. A minimal interpolation that respects both limits, the threshold and the energy distribution is the two-body phase-space form of~\cite{Khoze:2024hlb}, applied to the hadronic system with the energy it actually received and averaged over~\eqref{eq:dirichlet},
\begin{equation}
\br_{B+L}(E)\,\simeq\,\frac{1}{2}\,\Phi(E)\,,\qquad
\Phi(E)\,=\,\int_0^1dy\,\frac{y^{-1/4}(1-y)^{-3/4}}{\pi\sqrt2}\ \sqrt{1-\frac{m_p^2}{y^2E^2}}\ \theta(yE-m_p)\,,
\label{eq:interp}
\end{equation}
with $\br_{LR}=1-\br_{B+L}$. Near threshold $\Phi\simeq1.11\,(1-m_p/E)^{3/4}$, and at high energy $\Phi\simeq1-0.41\,(m_p/E)^{3/4}$. The result is shown in Figure~\ref{fig:br}, together with the two-body form~\eqref{eq:flipsig-e}. For example, $\br_{B+L}=0.26$ at $E=1.5\,\GeV$ and $0.43$ at $5\,\GeV$, against $0.39$ and $0.49$ for the two-body form. Equation~\eqref{eq:interp} is a model. The two limits, the threshold, the energy distribution and the slow $(m_p/E)^{3/4}$ approach to $1/2$ are not.

\subsection{The role of quantum fluctuations}
\label{sec:quantum}

The symmetry argument of section~\ref{sec:high} neglects the coupling of the string to the hard fronts. Its effect can be studied in the classical version of the bosonised model, four sine-Gordon fields on the half-line with the boundary condition~\eqref{eq:R}, which is a well-posed initial-value problem first solved numerically in~\cite{Dawson:1983cm}. We have repeated and extended these simulations. The details are given in appendix~\ref{sec:classical}, and here we state what they show.

The classical evolution confirms the ingredients of the picture. An incoming kink is reflected by the boundary into four half-kinks with the windings of the twisted state and with equal energies, as in~\eqref{eq:E4}. The string behind the fronts sits at the maximum of the potential and decays on the scale $1/m$. 
During the decay the energy of each sector changes only by an amount of order the gap, independent of $E$: by $1$--$3\,m_q$ for equal masses and by $6$--$7\,m_q$ in the light sector for hierarchical masses, where a fermion has the mass $8m_q$. This supports the assumption of section~\ref{sec:dirichlet} that the energies received by the sectors in the massless region remain their energy budgets during the decay.
With hierarchical masses the decay of the string in the heavy sectors sends a soft half-kink into the light sector, which makes its total charge integer, as described in section~\ref{sec:high}.

The classical evolution cannot give the branching ratios. It is deterministic, and without fluctuations the string always breaks the same way. For the mass patterns of the Standard Model the outcome is the helicity flip at every energy, and the $(B+L)$-violating channel appears only when the incoming flavour is heavier than the others, as found in~\cite{Dawson:1983cm}. The reason is the coupling to the hard fronts that was dropped in section~\ref{sec:high}, which pushes the string off the maximum of the potential towards one side by an amount of order $m/E$.

The equal weights~\eqref{eq:half} are therefore a quantum effect. The field on the string has vacuum fluctuations of order one, much larger than the classical bias at high energy, and it is these fluctuations that decide the direction in which the string breaks. In the linear phase of the instability the problem is Gaussian and can be solved. The result~\eqref{eq:gausscorr} is $\br_{B+L}\simeq \frac{1}{2}-(2\text{--}6)\,m/E$, where $m$ is the mass parameter of the classical model, in which a fermion has the mass $8m$.
 Whatever the precise coefficient, the bias vanishes as $m/E$ while the fluctuations do not depend on the energy, so~\eqref{eq:half} holds in the high-energy limit. The correction also falls faster with the energy than the kinematic one of section~\ref{sec:threshold}.

These estimates have their limits. The numbers that enter~\eqref{eq:gausscorr} are known up to a factor of two, the Gaussian treatment does not extend to the threshold region, where the hard-front approximation fails, and the classical model has no confinement. The classical model supports the picture and the size of the corrections, but it does not replace a full quantum computation of $\br_{B+L}(E)$.

\section{Electron- and proton-initiated processes}
\label{Sec:phys}

\subsection{An incoming electron}
\label{Sec:lepton}

Collecting~\eqref{eq:sigma0}, \eqref{eq:dirichlet}, \eqref{eq:channels}, \eqref{eq:half} and~\eqref{eq:interp}, an electron of energy $E\gg\Lqcd$ in the monopole rest frame undergoes exactly two $j_0$-wave processes on $M$, with their cross-sections shown on the right,
\begin{equation}
\begin{aligned}
e^-_R(E)+M&\,\to\, M\,+\,\big[\text{leptons}\big]_{L=1}(xE)\,+\,\big[\text{hadrons}\big]_{B=0,\,Q=0}(yE), &\quad \sigma&=\frac{\pi}{p^2}\Big[1-\frac{1}{2}\Phi(E)\Big],\\[2pt]
e^-_R(E)+M&\,\to\, M\,+\,\big[\text{leptons}\big]_{L=0}(xE)\,+\,\big[\text{hadrons}\big]_{B=-1,\,Q=-1}(yE), &\quad \sigma&=\frac{\pi}{p^2}\,\frac{1}{2}\Phi(E),
\end{aligned}
\label{eq:lepton_summary}
\end{equation}
where the subscripts give the lepton number $L$ of the leptonic system, and the baryon number $B$ and the total electric charge $Q$ of the hadronic system. The leptonic system is the content of the electron sector, with energy $xE$. It consists of electrons, positrons and photons, its electric charge is $-L$, and its content is given in~\eqref{eq:classes} below. The hadronic system has the energy $yE$, with $y=1-x$ and $x$ distributed as in~\eqref{eq:beta}, and its charge follows from the charge $-1$ of the incoming electron. The first line is the helicity flip. The second line is the $(B+L)$-violating process, the anti-nucleon synthesis. Its hadronic system contains an anti-proton with mesons of zero total charge, or an anti-neutron with mesons of total charge $-1$, as explained below~\eqref{eq:channels}. Both hadronic systems can contain further mesons and, at sufficiently high energy, baryon--anti-baryon pairs. For $E\ll\Lqcd$ only the first process remains, with no hadrons and $x=1$, as in~\eqref{eq:low}.

At high energy the energetic content of the electron sector is independent of the channel and is governed by its share $x$ of the energy (section~\ref{sec:hard}). Each of the two processes of~\eqref{eq:lepton_summary} can therefore be divided into two classes, according to whether the electron sector is hard or soft. The final states are
\begin{equation}
\begin{array}{lll}
\Delta (B+L)=0\,: & \text{(a)}\,\, e^-_L(xE)+[\text{hadrons}]_{B=0,\,Q=0}\,, & \text{(b)}\,\, e^-_{\rm soft}+[\text{hadrons}]_{B=0,\,Q=0}\,,\\[8pt]
\Delta (B+L)=-2\,: \quad & \text{(c)}\,\, e^-_L(xE)+e^+_{\rm soft}+\bar p(\bar n)+\text{mesons}\,, \quad & \text{(d)}\,\, \bar p(\bar n)+\text{mesons}+(e^+e^-)_{\rm soft}\,,
\end{array}
\label{eq:classes}
\end{equation}
where ``soft'' means an energy $\lesssim\Lqcd$, $(e^+e^-)_{\rm soft}$ denotes soft pairs or photons with zero lepton number. Further soft pairs can accompany every class, and $\bar p(\bar n)+\text{mesons}$ stands for $[\text{hadrons}]_{B=-1,\,Q=-1}$, in which the mesons have zero total charge for $\bar p$ and total charge $-1$ for $\bar n$. Classes (a) and (b) belong to the helicity flip, the first line of~\eqref{eq:lepton_summary}, and classes (c) and (d) to the anti-nucleon synthesis, the second line. They are not additional channels. There are still exactly two channels, and the division within each of them is by the share $x$, which varies continuously. Classes (a) and (c) occur when the electron sector has received a sizeable share of the energy, and classes (b) and (d) when $x$ is small and the whole sector is soft. Classes (a) and (d) are the familiar final states of the two processes. In (b) the electron number of the incoming electron is carried by a soft electron, and in (c) a hard helicity-flipped electron accompanies the anti-nucleon. In (d) the electron sector contains only soft pairs. At high energy the rate of each of the classes (a) and (c), with the electron above $cE$, is approximately one half of the first line of Table~\ref{Tab:hard}, which is $0.24$ for $c=0.1$. In the rest frame of the monopole the hard electron $e^-_L$ of classes (a) and (c) is emitted isotropically.

In~\cite{Candela:2025gwp} the anti-nucleon synthesis was considered for a relativistic monopole striking an atomic electron at rest, for which $E=\gamma_Mm_e$. The threshold $\gamma_M\gtrsim m_p/m_e\approx1.8\times10^3$ and the high-energy limit $\frac{1}{2}\times\pi/p^2$ of the cross-section are unchanged. What is new is that the limit is approached more slowly. Relative to the two-body
form~\eqref{eq:flipsig-e} used in~\cite{Candela:2025gwp}, the branching ratio is lower by a
factor of two or more just above the threshold, by about $10\%$ at $\gamma_M\simeq10^4$, and by
about $1\%$ at $\gamma_M\simeq2\times10^5$.

\medskip
\noindent{\it Energy spectra.}
The hadronic system receives the energy $yE$, with $y$ distributed as $g(y)$ of~\eqref{eq:beta}. How much of it becomes the kinetic energy $T_{\bar N}$ of the anti-nucleon depends on the hadronisation, which is not computed here. Two simple assumptions bracket the range. Either the anti-nucleon takes the whole available energy $yE-m_p$ (model A, the two-body limit), or it takes a uniformly distributed fraction of it (model B). For $E$ between $1.5$ and $10\,\GeV$ the mean kinetic energy rises from $0.45$ to $7.0\,\GeV$ in model A and is half of that in model B. The probability of $T_{\bar N}<0.1\,\GeV$ is at most $5\%$ in model A and falls from $0.29$ to $0.03$ in model B. The anti-nucleon is thus rarely produced at rest, and the annihilation signature of~\cite{Candela:2025gwp} is that of an anti-nucleon in flight.

The energy spectrum of the hard electron of classes (a) and (c) is $n_+(\omega|E)$ of~\eqref{eq:nspec}. The mean number of helicity-flipped electrons above $0.5\,\GeV$ per $j_0$-wave event rises from $0.28$ at $E=1.5\,\GeV$ to $0.57$ at $10\,\GeV$.

\subsection{A slow incoming proton}
\label{Sec:proton}

The processes~\eqref{eq:channels-u} and~\eqref{eq:channels-d} describe an incoming $s$-wave quark, and the symmetry argument of section~\ref{sec:high} applies to them verbatim. A quark is not an asymptotic state, however. The incoming object is a nucleon, the quark that reaches the core carries a fraction of its energy, and the spectator quarks take part in forming the final state.

As recalled in section~\ref{sec:kyg}, a non-relativistic proton with $\beta<0.2$ scatters on the monopole as a single object in the $j_0$-wave, and the overall factor is~\eqref{eq:sigma0} with $p=m_p\beta$. An $s$-wave quark that reaches the core, say $u_L^{\,1}$ with $E_q\sim m_p/3$, produces the twisted state of~\eqref{eq:callan-u}. With the spectator quarks, the two channels of~\eqref{eq:channels-u} become
\begin{equation}
p\,+\,M\,\to\, M\,+\,p\,+\,X_0\,,\qquad p\,+\,M\,\to\, M\,+\,\bar e\,+\,\text{mesons}\,.
\end{equation}
In the first channel the quark flips its helicity and the proton re-forms. A neutron would have to be accompanied by a positively charged meson, for example $n\,+\,\pi^+$, for which a slow proton does not have the energy. In the second channel the positron carries the electric charge of the proton and the mesons are neutral in total.

Let $p_q$ be the probability that a single encounter of a valence quark with the monopole core ends in the $(B+L)$-violating channel. It is the branching ratio $\br_{B+L}$ of section~\ref{Sec:br} for a quark of energy $E_q\sim\Lqcd$. At this energy every sector is in the strongly coupled regime, and $p_q$ is a number of order one that we cannot compute. The branching ratio of the catalysed decay of the proton is nevertheless close to one, as assumed in~\eqref{eq:flipsig-p}. The proton spends a time $R_p/\beta$ near the monopole, where $R_p\approx0.8\,$fm is the proton radius, while its valence quarks reach the core every $R_p/c$. A quark that only flips its helicity remains confined and the proton re-forms, so there are $n\sim1/\beta$ independent chances for the decay. The proton survives only if none of them succeeds, which has the probability $(1-p_q)^n$, so the probability of the decay is
\begin{equation}
P_{\rm decay}\,=\,1-(1-p_q)^n\,\simeq\,1-(1-p_q)^{1/\beta}\,\simeq\,1\qquad\text{for}\ \ p_q\gg\beta\,.
\label{eq:attempts}
\end{equation}
For the velocities relevant to catalysis searches, $\beta\sim10^{-3}$, this holds for any $p_q\gtrsim10^{-2}$. The counting treats the crossings as independent trials, which is a model of the bound-state dynamics and not a derivation.

\section{Conclusions}
\label{Sec:concl}

We have studied the evolution of the outgoing state of single-fermion scattering on the minimal monopole from the massless region near the core, where it is a single twisted state, to distances $r\gtrsim1/m$, where the fermion masses turn its topological line into an unstable string and the state decays into one of two physical channels. The results are listed in the introduction, and here we comment on their status. The overall cross-section~\eqref{eq:sigma0} and the energy distribution~\eqref{eq:dirichlet} are exact statements of the massless one-generation theory, and the two channels for each process~\eqref{eq:channels} in the massive theory follow from conservation laws. The equal weights~\eqref{eq:half} at high energy follow from the charge-conjugation symmetry of the decaying string, with a correction estimated in~\eqref{eq:gausscorr}, and the pure helicity flip~\eqref{eq:low} at low energy follows from decoupling. The interpolation~\eqref{eq:interp} between them and the counting~\eqref{eq:attempts} for a slow proton are models. Beyond the twisted-sector description, the analysis assumes that the sector energies of~\eqref{eq:dirichlet} label decohered branches, that the hard fronts do not influence how the string breaks, that the hard content of the electron sector does not depend on the channel, which holds for $xE\gg\Lqcd$, and that the one-generation boundary condition is the relevant one. We have also assumed that the monopole returns to its ground state. Dyonic bound states of the fermions on the monopole were studied in~\cite{Hook:2024vhf}, and whether such a state can form with the hierarchical masses of the Standard Model is left open here.

These results clarify the status of Callan's superposition, $\tfrac{1}{\sqrt2}\,|e_L\rangle + \tfrac{1}{\sqrt2}\,| (\overline{u})_{L1}\, (\overline{u})_{L2}\,(\overline{d}\,)_{R3}\rangle$ for an incoming electron. It does not describe the state at $r\ll1/m$. There the outgoing state is the twisted state, whose flavour numbers are sharp and equal to $\pm\frac{1}{2}$, while a superposition of two integer states would have the same mean values but fluctuating charges. At $r\gtrsim1/m$, after the string has broken, the final state is a superposition of the two channels of~\eqref{eq:channels}. It differs from Callan's form in two ways. The weights are the energy-dependent branching ratios, equal only for $E\gg m_p$, and each component is an inclusive state dressed by soft secondaries and neutral radiation. The two components differ in baryon and lepton number, which are conserved away from the monopole. Their relative phase is therefore not observable, and the superposition is equivalent to a statistical mixture. In this sense Callan's statistical interpretation is recovered for the asymptotic state.

For phenomenology, the electron-initiated processes are collected in~\eqref{eq:lepton_summary}. Relative to~\cite{Candela:2025gwp}, the threshold of anti-nucleon synthesis is confirmed, the hadronic system typically carries most of the energy, the asymptotic rate is half the unitarity bound and is approached slowly, and a hard helicity-flipped electron accompanying the anti-nucleon is a new signature. The equal weights are specific to the twist charge $\frac{1}{2}$ of the one-generation theory on $M$. For the higher monopoles of~\cite{Khoze:2024hlb} the two channels are not related by a symmetry, and their branching ratios are left for future work. For a slow proton the catalysed decay has a probability close to one, as assumed in~\eqref{eq:flipsig-p}.

Two things remain open in the one-generation theory. They are the shape of $\br_{B+L}(E)$ through the threshold region and the hadronisation of the quark sectors, which decides how often the anti-nucleon carries the hadronic energy. Both are obscured in QCD by confinement. They could be studied without it in the Georgi--Glashow $SU(2)$ model of~\cite{Khoze:2023kiu} with $N_f=4$ Weyl doublets and two gauge-invariant masses $m_\ell\ll m_q$, which has the same $s$-wave boundary condition, twisted sector and pair of channels as $M$. We leave this for future work, together with the extension to three generations, which depends on how the boundary condition at the monopole core is formulated for several generations, for all flavours at once or generation by generation.

\section*{Acknowledgements}

I am grateful to David Tong for sending me Ref.~\cite{Tong:2026apr}, which was posted on the arXiv on the same day as the first version of the present paper.
This work was supported by the STFC under IPPP Consolidated and National Centre grants UKRI4227 and UKRI4534.
 The AI system Claude (Anthropic) was used for proof-reading and editing the text, for making the figures, and for writing the code for the numerical computations used in sections~\ref{sec:dirichlet} and~\ref{sec:hard} and in appendices~\ref{app:hard} and~\ref{sec:classical}. All physics arguments and results were established and verified by the author, who takes full responsibility for the content.

\appendix
\section{The symmetry argument in bosonised form}
\label{app:reflection}

In the bosonised description of section~\ref{sec:twisted} the charge conjugation used in section~\ref{sec:high} is a reflection of the soft fields, and the argument for~\eqref{eq:half} can be followed explicitly. The four fields share one boundary value $\phi_b(t)$ at the monopole and roll in the same direction there, which is the bosonised form of the selection rule of section~\ref{sec:selrule}. The direction of the roll is the sign of the charge absorbed by the monopole.

\begin{figure}[ht]
\centering
\begin{tikzpicture}[>=stealth,scale=1.0,font=\footnotesize,
  declare function={sb(\x)=0.5*exp(-0.05*\x)*cos((0.75*\x) r);
                    sa(\x)=-0.32*exp(-0.35*(\x-5))*cos((2.2*(\x-5)) r);}]
\draw[->,gray] (0,-0.2) -- (7.9,-0.2) node[right,black] {$r$};
\draw[->,gray] (0,-0.5) -- (0,4.1) node[above,black] {$\phi_j(r)$};
\node[below] at (0,-0.5) {$M$};
\draw[gray,densely dotted] (0,0.6) -- (7.6,0.6);
\draw[gray,densely dotted] (0,2.2) -- (5.0,2.2);
\draw[gray] (-0.08,0.6) -- (0.08,0.6);
\draw[gray] (-0.08,2.2) -- (0.08,2.2);
\draw[gray] (-0.08,3.8) -- (0.08,3.8);
\node[left] at (-0.1,0.6) {$0$};
\node[left] at (-0.1,2.2) {$\pi$};
\node[left] at (-0.1,3.8) {$2\pi$};
\draw[thick,black!50,densely dashed,domain=0:5.0,samples=90,smooth] plot (\x,{2.2-sb(\x)});
\draw[thick,black!50,densely dashed] (5.0,{2.2-sb(5.0)}) -- (5.0,{0.6-sa(5.0)});
\draw[thick,black!50,densely dashed,domain=5.0:7.6,samples=60,smooth] plot (\x,{0.6-sa(\x)});
\draw[very thick,domain=0:5.0,samples=90,smooth] plot (\x,{2.2+sb(\x)});
\draw[very thick] (5.0,{2.2+sb(5.0)}) -- (5.0,{0.6+sa(5.0)});
\draw[very thick,domain=5.0:7.6,samples=60,smooth] plot (\x,{0.6+sa(\x)});
\draw[<->] (0.8,2.2) -- (0.8,{2.2+sb(0.8)});
\node[left] at (0.8,2.4) {$\delta_j$};
\draw[<->] (6.43,0.6) -- (6.43,{0.6+sa(6.43)});
\node[above] at (6.43,0.85) {$\varepsilon_j$};
\draw[gray] (5.0,-0.3) -- (5.0,-0.1);
\node[below] at (5.0,-0.3) {hard front};
\node[right] at (5.1,1.45) {fixed step $\pi$};
\fill (0,{2.2+sb(0)}) circle (1.8pt);
\node[above right] at (0.0,{2.25+sb(0)}) {$\phi_b$};
\end{tikzpicture}
\caption{The soft field in the flavour sector $j$ at a fixed time, shown for the sign $+\pi$. The hard front enters as a fixed step of height $\pi$. Behind it the field is $\phi_j=\pi+\delta_j$, a fluctuation $\delta_j$ around the maximum of the cosine potential, and ahead of it $\phi_j=\varepsilon_j$, a fluctuation around the vacuum. The four sectors share the boundary value $\phi_b$ at the monopole. The dashed curve is the configuration obtained by the reflection $\delta_j\to-\delta_j$, $\varepsilon_j\to-\varepsilon_j$, which leaves the step unchanged. A configuration that rolls towards $2\pi$ at the boundary is mapped into one that rolls towards $0$.}
\label{fig:delta}
\end{figure}
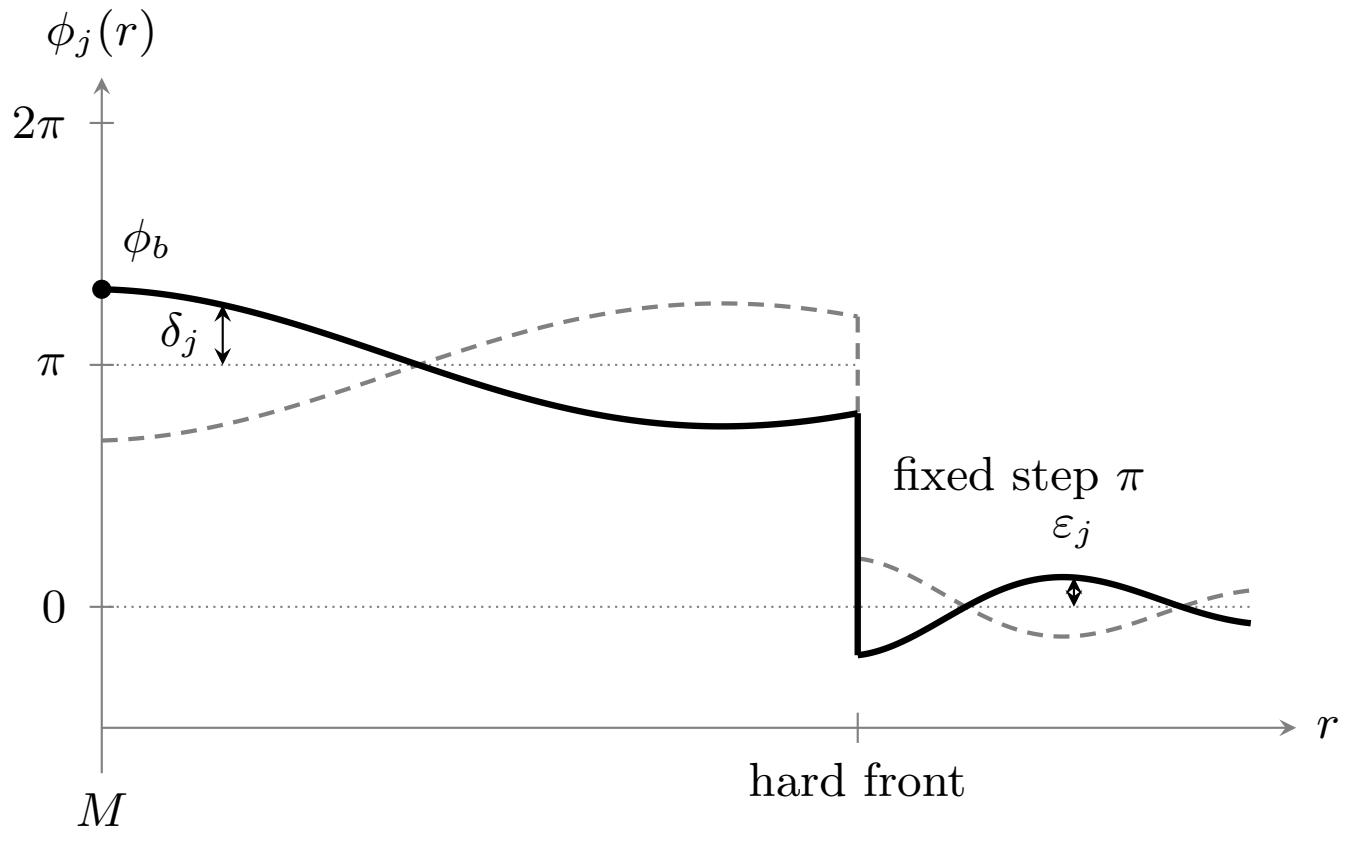

At high energy the hard front of flavour $j$ acts on the soft modes only as a fixed topological step of $\pm\pi$ in $\phi_j$. Behind the front we write $\phi_j=\pm\pi+\delta_j$, with $\delta_j=0$ initially, and ahead of it $\phi_j=\varepsilon_j$, a fluctuation around the vacuum, as shown in Figure~\ref{fig:delta}. The simultaneous reflection $\delta_j\to-\delta_j$, $\varepsilon_j\to-\varepsilon_j$ in all four sectors is an exact symmetry of this soft-field problem as a whole, that is, of its bulk equations of motion, of the boundary condition at the monopole, of the matching across the fronts and of the initial state, for any values of the masses. In the bulk the kinetic terms are quadratic and the potentials are even, $-\cos(\pm\pi+\delta_j)=\cos\delta_j$ behind the front and $-\cos\varepsilon_j$ ahead of it. The boundary condition~\eqref{eq:R} is linear and homogeneous in the fields. The fixed step requires $\delta_j=\varepsilon_j$ at the front, which the reflection preserves. The initial state, the vacuum fluctuations around $\delta_j=\varepsilon_j=0$, is even. The reflection exchanges the two directions of the common boundary roll, $\phi_b\to0$ and $\phi_b\to2\pi$, that is, the helicity flip and the $(B+L)$-violating channel, which therefore have equal probability.

Since $\phi_j$ is defined modulo $2\pi$, the reflection is $\phi_j\to-\phi_j$ on both sides of the front. It is the charge conjugation of section~\ref{sec:high}, acting on the soft modes with the hard fronts held fixed. The potential behind the front is even in $\delta_j$ only if the field sits exactly at the maximum of the cosine potential, which is the requirement that the twist charge is $\frac{1}{2}$. The potential of a massless sector is trivially even, so the hierarchy of the masses does not affect the argument.

\section{The fermionic content of a hard front}
\label{app:hard}

This appendix gives the computation behind section~\ref{sec:hard}, that is, the origin of~\eqref{eq:fourpt}, the exact occupation numbers whose continuum limit is the spectrum~\eqref{eq:nspec}, and the entries of Table~\ref{Tab:hard}.

\medskip
\noindent{\it The question.}
In bosonised form the outgoing state is a product, over the four sectors, of vertex operators with the charges $R_{j1}=\pm\frac{1}{2}$. This fixes the charge of each sector and the distribution of its energy (section~\ref{Sec:massless}), but it does not say what the sector contains in terms of fermions, which is what a detector registers. A vertex operator of charge one creates a single fermion. 
A vertex operator of charge $\frac{1}{2}$ creates a state without a definite number of fermions, and the question is how many fermions and anti-fermions of a given energy this state contains.
In the massless region the sectors are decoupled, so this is a question about a single free chiral Dirac fermion $f$ and the state created by $V=\,:\!e^{i\phi/2}\!:$, in the normalisation in which $f^\dagger=\,:\!e^{i\phi}\!:$. For a sector of charge $-\frac{1}{2}$ the roles of $f$ and of its anti-particle are exchanged.

\medskip
\noindent{\it Set-up.}
We use the box normalisation of section~\ref{sec:dirichlet}, in which the momenta are multiples of $2\pi/L$, and measure energies in this unit, so that they are non-negative integers. In the radial variable $z$ of the chiral theory an operator at the origin creates a state, and moving it to the radius $\xi<1$ multiplies the component of energy $n$ by $\xi^n$. The variable $\xi$ is thus the counting variable of section~\ref{sec:dirichlet}. The norm of the state created by $V$ is resolved in energy by
\begin{equation}
\big\langle V^\dagger(1)\,V(\xi)\big\rangle\,=\,(1-\xi)^{-1/4}\,=\,\sum_{n\geq0}w_n\,\xi^n\,,\qquad w_n\,=\,\frac{\Gamma(n+\frac{1}{4})}{\Gamma(\frac{1}{4})\,n!}\,,
\label{eq:wn}
\end{equation}
where $w_n$ is the weight of the component in which the sector has the energy $n$. This is the generating function of section~\ref{sec:dirichlet} with $R_{j1}^2=\frac{1}{4}$. The product over the four sectors is $(1-\xi)^{-1}$, in which every total energy $E$ has weight one, as it must for an incoming fermion of definite energy. The other three sectors together have the generating function $(1-\xi)^{-3/4}$, and the resulting distribution of the energy of one sector is the discrete form of the Beta distribution~\eqref{eq:beta} of the lepton share.

\medskip
\noindent{\it The four-point function.}
The occupation numbers of the fermion modes in the state created by $V$ are contained in the correlator of $f^\dagger f$ with $V$ and its conjugate. For vertex operators of a free chiral boson with charges $q_i$ at the points $z_i$ the correlator is $\prod_{i<j}(z_i-z_j)^{q_iq_j}$~\cite{Ginsparg:1988ui,DiFrancesco:1997nk}, and with the charges $(-\frac{1}{2},+1,-1,+\frac{1}{2})$ at the points $(1,z',z,\xi)$ this is~\eqref{eq:fourpt},
\begin{equation}
\big\langle V^\dagger(1)\,f^\dagger(z')\,f(z)\,V(\xi)\big\rangle\,=\,\frac{(1-\xi)^{-1/4}}{z'-z}\left[\frac{(1-z)\,(z'-\xi)}{(1-z')\,(z-\xi)}\right]^{1/2}\,.
\nonumber
\end{equation}
In the region $\xi<|z|<|z'|<1$ we expand $1/(z'-z)=\sum_{n\geq0}z^n z'^{-n-1}$ and write the square root as $(z'/z)^{1/2}A(z)B(z')$, with
\begin{equation}
\begin{aligned}
A(z)&\,=\,(1-z)^{1/2}\Big(1-\frac{\xi}{z}\Big)^{-1/2}\,=\,\sum_pA_p(\xi)\,z^p\,,\\
B(z')&\,=\,(1-z')^{-1/2}\Big(1-\frac{\xi}{z'}\Big)^{1/2}\,=\,\sum_qB_q(\xi)\,z'^{\,q}\,.
\end{aligned}
\label{eq:AB}
\end{equation}
The coefficients are power series in $\xi$ built from the binomial coefficients of $(1-u)^{\pm1/2}$. The half-integer powers of $z$ and $z'$ show that the fermion has integer modes in the presence of $V$, which is the twisted sector. The coefficient of $z^{m-1/2}z'^{-m-1/2}$ in the correlator is $(1-\xi)^{-1/4}\,C_m(\xi)$ with
\begin{equation}
C_m(\xi)\,=\,\sum_{p\leq m}A_p(\xi)\,B_{-p}(\xi)\,.
\label{eq:Cm}
\end{equation}
For $m=-k<0$ it counts the quanta of $f$ of energy $k$, and for $m=k>0$ it is one minus the number of anti-particles of energy $k$. Without the vertex operators $A=B=1$, so that $C_m=1$ for $m\geq0$ and $C_m=0$ for $m<0$, the empty vacuum. The power of $\xi$ resolves the occupation numbers by the energy of the sector. To obtain the state of total energy $E$ we multiply by the generating function $(1-\xi)^{-3/4}$ of the other three sectors and take the coefficient of $\xi^E$, which is normalised to one. The mean numbers of quanta of $f$ and of its anti-particle with energy $k$ are therefore
\begin{equation}
N_+(k|E)\,=\,\big[\xi^E\big]\,\frac{C_{-k}(\xi)}{1-\xi}\,,\qquad
N_-(k|E)\,=\,1-\big[\xi^E\big]\,\frac{C_{k}(\xi)}{1-\xi}\,,
\label{eq:Npm}
\end{equation}
where $[\xi^E]$ denotes the coefficient of $\xi^E$. These expressions are exact at any $E$.

\medskip
\noindent{\it The continuum limit and a numerical check.}
The continuum limit is $E\to\infty$ at fixed $k/E$. In this limit the occupation numbers~\eqref{eq:Npm} should reduce to the spectrum~\eqref{eq:nspec} of~\cite{Tong:2026apr}, with $\omega=k$. We have not derived this limit analytically, and have checked it numerically instead. We have evaluated~\eqref{eq:Npm} exactly at $E=400$, as power series in $\xi$. The energy sum rule $\sum_kk\,(N_++N_-)=E/4$ holds to machine precision. The charge is $\sum_k(N_+-N_-)=0.498$, and it is exactly $\frac{1}{2}$ when the zero mode $k=0$ is included. The comparison with~\eqref{eq:nspec} at a few energies is
\begin{equation} \nonumber
\begin{array}{l|cccc}
k/E &\ 0.05\ &\ 0.25\ &\ 0.5\ &\ 0.9\
\\\hline
E\,N_+(k|E)\text{ at }E=400 & 2.480 & 0.718 & 0.464 & 0.336
\\
E\,n_+(\omega|E) & 2.480 & 0.718 & 0.464 & 0.336
\\\hline
E\,N_-(k|E)\text{ at }E=400 & 1.592 & 0.151 & 0.027 & 0.0005
\\
E\,n_-(\omega|E) & 1.592 & 0.151 & 0.027 & 0.0005
\end{array}
\end{equation}
The two agree to better than one part in a thousand for $N_+$, and to the accuracy shown for $N_-$.

\medskip
\noindent{\it The table.}
The first two lines of Table~\ref{Tab:hard} are the integrals of the spectrum~\eqref{eq:nspec} from $cE$ to $E$, computed by numerical quadrature. The exact sums of~\eqref{eq:Npm} at $E=400$ agree with them to within $0.003$. The last line is the integral from $c$ to $1$ of the Beta density~\eqref{eq:beta} of the lepton share. The first two lines are mean numbers of quanta. For $c\geq\frac{1}{2}$ energy conservation allows at most one quantum above $cE$ in the whole state, so they are probabilities, and a quantum above $cE$ in a sector implies that the share of this sector exceeds $c$. The ratio of the first line to the last is then the conditional probability quoted in section~\ref{sec:hard}, $0.188/0.219=86\%$ at $c=\frac{1}{2}$. The computation gives the mean occupation numbers at fixed total energy. It does not give the joint distribution of the energy of the hard fermion and of the share of its sector, or the correlations between the quanta, which would require higher correlators.

\section{The classical bosonised model}
\label{sec:classical}

The mechanism of sections~\ref{Sec:mass} and~\ref{Sec:br} can be tested in the classical version of the bosonised model, which is a well-posed initial-value problem. It has four fields on the half-line with ${\cal L}=\sum_j[\frac{1}{2}(\partial\phi_j)^2-m_j^2(1-\cos\phi_j)]$, which is~\eqref{eq:bosonisation} with~\eqref{eq:mass} rescaled by $4\pi$, with $m_j^2$ standing for $4m_j\Lambda$. A fermion is a $2\pi$ kink of mass $8m_j$, and the boundary condition is~\eqref{eq:R}. The problem was first studied numerically in~\cite{Dawson:1983cm}. For a single incoming soliton and equal masses the half-solitons were found there to recombine into the helicity-flipped soliton, and the three-soliton state appeared only when the incoming flavour was heavier than the others. We have repeated the computation with~\eqref{eq:R} imposed exactly, for an incoming kink in sector 1 of energy $E=8m_1\gamma$, with $\gamma=5$--$40$ for equal masses, and with $\gamma=100$ and $200$, that is $E=40$ and $80\,m_q$, for the light incoming flavour. We considered equal masses, the hierarchies $m_1=0.05\,m_q$ (incoming electron) and $m_4=0.05\,m_q$ (incoming quark), and a heavy incoming flavour. The classical theory cannot give the quantum branching ratios, but it tests the ingredients of the argument.

\begin{itemize}
\item {\it Reflection and energy partition.} On impact $\phi_b$ rises from $0$ to $\pi$, and the outgoing fronts carry the windings $(+\frac{1}{2},-\frac{1}{2},-\frac{1}{2},-\frac{1}{2})$. Just after the reflection the energies of the four sectors are $(40.2,39.9,39.9,39.9)\,m_q$ at $\gamma=20$, against $E/4=40\,m_q$. This is the equipartition~\eqref{eq:E4} with the expected $O(m/E)$ correction.
\item {\it Instability.} The region behind the fronts sits at the maximum of the potential and decays on the scale $1/m$. At $\gamma=20$ the boundary value is $\phi_b/\pi=0.95,\,0.87,\,0.70,\,0.30$ at $t=2,3,4,5$ in units of $1/m_q$ after impact.
\item {\it Energy budget.} During the decay the sector energies change only by amounts of order the gap, independent of $E$. With equal masses the boundary moves $1.0\,m_q$ out of each of three sectors into the fourth, and the energies become $(43.1,38.9,38.9,38.9)\,m_q$ at $\gamma=20$. With $m_1=0.05\,m_q$ the electron sector gains $6$--$7\,m_q$ at both $E=40\,m_q$ and $80\,m_q$. This is the soft half-kink emitted from the boundary into the electron sector, described in section~\ref{sec:high}. The partition~\eqref{eq:dirichlet} thus fixes the budget of each sector up to a shift of order the gap, as assumed in section~\ref{sec:dirichlet}.
\item {\it A deterministic choice.} Without fluctuations the roll always goes the same way. For the mass patterns of the Standard Model (equal masses, a light incoming flavour, a light outgoing flavour) it is the helicity flip, $\phi_b\to0$, at every energy. A heavy incoming flavour, $m_1\geq1.75\,m_q$, gives the $(B+L)$-violating channel, and the switch occurs between $m_1/m_q=1.5$ and $1.75$ independently of $\gamma$. This is the mass dependence found in~\cite{Dawson:1983cm}.
\end{itemize}

The classical bias has a simple origin, the coupling to the hard fronts that was dropped in section~\ref{sec:high}. Each front is a light-like source $-m_j^2\sin\phi_j$ for the soft field. In the three sectors that received an anti-half-kink it pushes the common boundary value back towards $0$, the helicity flip. In the sector of the incoming flavour it pushes towards $2\pi$, the $(B+L)$-violating channel. The net seed is proportional to $\big(\sum_{j\neq1}m_j^2-m_1^2\big)/E$. It is of order $m/E$, and it changes sign at $m_1=\sqrt3\,m_q$ for three equal quark masses, as observed. From the growth of the unstable mode in the runs, its initial amplitude is
\begin{equation}
b_0\,\simeq\,(1.2\text{--}1.8)/\gamma\,\simeq\,(10\text{--}14)\,m/E\,.
\label{eq:b0}
\end{equation}

As stated in section~\ref{sec:quantum}, the equal weights~\eqref{eq:half} require fluctuations of the soft field at the maximum of the potential that exceed this bias. The decision can be computed where it is made, in the linear phase of the instability. Writing $\phi_j=\pi+\delta_j$, the potential near its maximum is an inverted parabola, $V\simeq2m^2-\frac{1}{2}m^2\delta_j^2$, and as long as $\delta_j$ is small the equation of motion is linear. This is the linear phase. In it the sum field near the boundary obeys $\ddot\delta_k=(m^2-k^2)\delta_k$, so the modes with $k<m$ grow exponentially, and for a quadratic Hamiltonian the Wigner function evolves by the classical flow. The amplitude of the growing mode is then a Gaussian variable, with mean $b_0$ and with a variance $\sigma^2$ given by the vacuum fluctuations of the sum field in the unstable band. Once the amplitude is of order one the linear phase ends, and the mode rolls classically to the well on its side. The sign of the amplitude at that point therefore decides the channel. Hence
\begin{equation}
\br_{B+L}\,=\,\frac{1}{2}\,{\rm erfc}\Big(\frac{b_0}{\sqrt2\,\sigma}\Big)\,\simeq\,\frac{1}{2}-\frac{b_0}{\sqrt{2\pi}\,\sigma}\qquad(E\gg m)\,.
\label{eq:gauss}
\end{equation}
At the free-fermion point $\hbar_{\rm eff}=4\pi$ in the normalisation of this section, and for the massless vacuum at the passage of the front $\sigma^2(t)=2\ln(mt)+1$, which gives $\sigma=1.0$--$1.8$ for $t=1$--$3$ in units of $1/m$. The range of $\sigma$ reflects the time at which the linear phase is taken to end. This time is not sharply defined, because at the free-fermion point the fluctuations are of order one from the start and the linear phase is short. The fluctuation exceeds the classical seed by a factor of $3$--$8$ at $\gamma=5$ and $10$--$30$ at $\gamma=20$, and
\begin{equation}
\br_{B+L}\,\simeq\,\frac{1}{2}-(2\text{--}6)\,\frac{m}{E}\,.
\label{eq:gausscorr}
\end{equation}
The correction is $5$--$15\%$ at $E=40\,m$ and below $4\%$ for $E\gtrsim160\,m$. It falls faster with the energy than the kinematic one of~\eqref{eq:interp}, which decreases only as $(m_p/E)^{3/4}$. Equation~\eqref{eq:gauss} is exact for the linear phase, but the values of $b_0$ and $\sigma$ are estimates at the factor-of-two level, and the result does not extend to the threshold region $E\sim m$, where the hard-front approximation fails.

\bibliographystyle{JHEP}
\bibliography{Massive-fermi-Mono}

\end{document}